\documentclass[12pt]{article}
\usepackage{amsmath,amssymb,bm,amsthm}
\usepackage{bbm}
\usepackage{booktabs,threeparttable,array,tabularx}
\usepackage{float}
\usepackage{microtype}
\usepackage[T1]{fontenc}
\usepackage{lmodern}
\usepackage{graphicx}
\usepackage[font=small,labelfont=bf,tableposition=top]{caption}
\usepackage{geometry}
\usepackage{setspace}
\usepackage{natbib}
\usepackage{placeins}
\usepackage{xr}
\usepackage[final]{hyperref}
\hypersetup{colorlinks=true,linkcolor=blue,citecolor=blue,
  filecolor=magenta,urlcolor=blue}
\allowdisplaybreaks
\theoremstyle{plain}
\newtheorem{theorem}{Theorem}

\newtheorem{proposition}{Proposition}

\newtheorem{definition}{Definition}
\theoremstyle{definition}
\newtheorem{assumption}{Assumption}
\newtheorem{remark}{Remark}
\renewenvironment{itemize}{\begin{enumerate}}{\end{enumerate}}

\begin{document}
\title{Robust Inference for Spiked Eigenspace
under Elliptical Distributions}
\date{}
\author{Zeyu Wang\thanks{Email: \href{mailto:wanzeyu0013@mail.ustc.edu.cn}{wanzeyu0013@mail.ustc.edu.cn}}
\quad
Qing Yang\thanks{Email: \href{mailto:yangq@ustc.edu.cn}{yangq@ustc.edu.cn}}
\quad
Xiao Han\thanks{Email: \href{mailto:xhan011@ustc.edu.cn}{xhan011@ustc.edu.cn}}\\[0.5em]
{\small Faculty of Business for Science \& Technology, School of Management,}\\
{\small University of Science and Technology of China, Hefei, China}
}
\maketitle

\begin{abstract}
In this paper, we study two eigenspace testing problems under
high-dimensional elliptical spiked models: whether the spiked
eigenspace equals a given subspace, and whether a given vector
is a spike eigenvector. We propose a \textbf{Robust Functional
Projection} (RFP) test and a \textbf{Robust Spike Direction}
(RSD) test for these two problems, respectively. Based on the
spatial-sign covariance matrix $\bm C_n$, both tests achieve
robustness by remaining invariant to the radial distribution
without moment conditions and allowing the background level
to be unknown. To construct and analyze both tests, we establish
central limit theorems for \textbf{G}eneralized \textbf{E}lliptical \textbf{S}patial-sign \textbf{S}pectral statistics (GESS)
$\operatorname{tr}\{f(\bm C_n)\bm B_n\}$.
The results cover a general class of deterministic weighting
matrices $\bm B_n$ whose ranks are either comparable to the
dimension or of smaller order, providing a common theoretical
foundation for both tests. The central idea in constructing
both tests is to avoid the need for consistent estimations of
individual spike eigenvalues by working directly with
GESS. Simulations illustrate the finite-sample performance of the
procedures, and an analysis of LOCATA acoustic data demonstrates
their use in assessing a signal subspace and a steering direction,
both constructed from measured source and array geometry.

\end{abstract}

\section{\textbf{Introduction}}\label{Introductio}
Microphone arrays collect multichannel acoustic observations whose
spatial structure is central to sound source localization and
separation. A key object is the signal subspace, which represents
source-related variation across the microphone channels
\citep{nakadai2018music,evers2020locata}.
Acoustic observations can exhibit heavy tails and substantial
variation in signal power, motivating distributional models beyond
the Gaussian setting. In particular, multivariate complex Student's
$t$ distributions and elliptically contoured stable distributions
have been used to model multichannel audio observations
\citep{kitamura2016student,leglaive2017alpha}.
These developments motivate studying signal subspaces within the
broader class of elliptical distributions.
We consider observations of the form
\begin{equation}\label{gabaji}
\bm x_i=\zeta_i\bm T^{1/2}\bm u_i,
\qquad i=1,\ldots,n,
\end{equation}
where $\zeta _i>0$ is a radial variable, $\bm u_i$ is uniformly
distributed on the unit sphere in $\mathbb R^p$ and independent
of $\zeta_i$, and $\bm T$ is a positive definite scatter matrix. When $\mathbb{E}(\zeta_i^2/p)<\infty$, the covariance matrix is
a positive scalar multiple of the scatter matrix:
$\operatorname{Cov}(\bm x_i  )
=\{\mathbb{E}(\zeta_i^2)/p\}\bm T$.
The scatter matrix remains well defined even when the
covariance matrix does not exist. We use real-valued notation
throughout the paper, while in the acoustic application, complex frequency-domain
observations are represented by stacking their real and imaginary
parts.

A low-rank signal component together with a spherical noise
background forms the familiar spiked spectral structure which has been extensively studied in random matrix
theory
\citep{johnstone2001distribution,baik2005phase,
baik2006eigenvalues,paul2007asymptotics,bai2008central,
benaych2011eigenvalues,bai2012sample,
bloemendal2016principal,wang2017asymptotics,
johnstone2018notes,cai2020limiting,jiang2021generalized,
bao2022statistical,zhang2022asymptotic,fan2022asymptotic}.
More recently, \citet{xie2026limiting} studied spiked models
under elliptical distributions. Such a
background also arises after whitening by the population noise
covariance in array-processing models
\citep{nadakuditi2010fundamental}.
Motivated by this structure, we consider an elliptical model with
\begin{equation}
\bm T
=
\sigma^2\bm I_p+\bm L\bm L^{\prime}
=
\sigma^2\bm I_p+
\sum_{j=1}^{r_n}d_j\bm v_j\bm v_j^{\prime},
\qquad d_j>0,\quad \sigma^2>0,
\label{intro:spiked-scatter}
\end{equation}
where $\bm L$ has full column rank $r_n<p$ and
$\bm v_1,\ldots,\bm v_{r_n}$ are orthonormal.
We refer to the common non-spiked eigenvalue $\sigma^2$
as the \textbf{background level}. Here the background level $\sigma^2$ is unknown.
The signal subspace and its orthogonal projection are
\begin{equation}
\operatorname{col}(\bm L)
=
\operatorname{span}\{\bm v_1,\ldots,\bm v_{r_n}\},
\qquad
\mathcal Z_n
=
\sum_{j=1}^{r_n}\bm v_j\bm v_j^{\prime}.
\label{intro:signal-space}
\end{equation}
Thus, under this model, the signal subspace is precisely the
spiked eigenspace of $\bm T$.
Physical information can provide a candidate for this space.
Measured source positions, together with the known microphone
geometry and a propagation model, determine nominal array responses
whose span specifies a candidate signal subspace
\citep{nakadai2018music}.
For example, the LOCATA corpus provides acoustic recordings
together with source and array positions obtained using an optical
tracking system \citep{evers2020locata}.
These measurements allow a candidate space to be constructed from
the recorded geometry. We denote its orthogonal projection by $\mathcal Z_0$.

However, the resulting candidate subspace need not coincide with the signal subspace of the received observations.
Reverberation, imperfect array calibration, and errors in the
assumed propagation model can lead to discrepancies between
the assumed and actual array responses.
Steering-vector mismatch and its consequences have  also been studied
extensively in robust beamforming
\citep{vorobyov2003robust,li2003robust,lorenz2005robust,
kim2008robust,gu2012robust,vorobyov2013principles}.
For microphone arrays, sensitivity to errors in the assumed
array characteristics has also motivated robust beamformer
designs \citep{doclo2003design,doclo2007superdirective}.
In acoustic source localization, \citet{nakadai2018music}
report discrepancies between geometrically calculated
steering vectors and the actual array responses.
Consequently, accurate position measurements alone do not
guarantee that the resulting nominal space describes the
observed signal subspace.
This motivates testing
\begin{equation}
H_0:\mathcal Z_n=\mathcal Z_0,
\qquad\text{versus}\qquad
H_1:\mathcal Z_n\ne\mathcal Z_0,
\label{intro:eigenspace-test}
\end{equation}
when the dimension $p$ and sample size $n$ are of the same order.
The purpose is to assess whether a physically specified signal
space is same with the population spiked eigenspace.

The difficulty already arises in the classical spiked covariance model,
when $p/n\to c\in(0,\infty)$ and the population spikes remain bounded. The
leading sample eigenvectors are generally not consistent estimators of their
population counterparts, even in the Gaussian case
\citep{paul2007asymptotics,johnstone2009consistency}. Consequently, even under $H_0$, the estimated spiked eigenspace
may fail to converge to $\bm{\mathcal Z}_0$. We therefore need to seek a statistic that directly tests the specified
eigenspace in high dimensions, without estimating the
population eigenspace from sample eigenvectors.
\citet{hu2024generalized} address this difficulty through functional
projection (FP).  
However, FP is derived under the independent-component model
\[
    \bm x_i=\bm T^{1/2}\bm z_i,
    \qquad
    \bm z_i=(z_{i1},\ldots,z_{ip})^{\prime},
\]
where the entries of $\bm z_i$ are independent and identically distributed
with zero mean and unit variance and this coordinate independence is fundamental
to their methodology and theory. The spiked population covariance matrix is
normalized to have an identity background, so a general background level
$\sigma^2$ must be given before the procedure can be applied. Moreover, the
entries of $\bm z_i$ are required to have a finite fourth moment, and the value
of this moment enters explicitly into the asymptotic mean and variance.
Consequently, the practical construction of FP requires both the background
level and the fourth moment to be known. As a result, FP is less readily implementable when either
the background level or the fourth moment is unknown.

The moment requirement above is closely tied to the use of the sample
covariance matrix, which retains the information of the norms of the observations.
Spatial signs provide a natural way to remove these norms while
preserving the population eigenspace of interest under elliptical distributions. Recalling the definition of elliptical distribution in (\ref{gabaji}), we obtain the spatial-sign normalization as
\[
    \frac{\bm x_i}{\|\bm x_i\|}
    =
    \frac{\bm T^{1/2}\bm u_i}
    {(\bm u_i^{\prime}\bm T\bm u_i)^{1/2}},
\]
whenever $\bm x_i\neq\bm0$, so the radial variable $\zeta_i$
disappears completely.
Define
\[
    \bm y_i=\frac{\sqrt p\,\bm x_i}{\|\bm x_i\|},
    \qquad
    \bm C_n=\frac{1}{n}\sum_{i=1}^n
    \bm y_i\bm y_i^{\prime},
    \qquad
    \bm\Sigma=\mathbb E(\bm y_i\bm y_i^{\prime}).
\]
We refer to $\bm C_n$ and $\bm\Sigma$ as the sample and population
\textbf{S}patial-\textbf{s}ign \textbf{C}ovariance \textbf{M}atrices (SSCM), respectively. Under the elliptical
model, $\bm\Sigma$ and $\bm T$ share the same eigenvectors and 
the same eigenspaces \citep{cheng2019testing}. 
Their eigenvalues generally differ. But after trace normalization, the eigenvalues of the population SSCM are in one‑to‑one correspondence with those of the scatter matrix \citep{durre2016eigenvalues}. Consequently, the population SSCM and $\bm T$ share the same
spiked eigenspace, which also coincides with that of the
covariance matrix whenever the latter exists. The hypothesis
in \eqref{intro:eigenspace-test} can therefore be formulated equivalently
in terms of $\bm\Sigma$.

Spatial-sign normalization removes the radial variable from
$\bm y_i$, so the distribution of the SSCM is invariant to
the radial distribution and no moment conditions on the
radial variable are required.
This invariance provides robustness under heavy-tailed
elliptical distributions, hence spatial signs have been used for robust principal
component analysis, correlation estimation, high-dimensional
sphericity testing, covariance testing and spectral inference
\citep{taskinen2012robustifying,durre2015spatial,
zou2014multivariate,cheng2019testing,li2022eigenvalues}.

Building on these properties, we propose \textbf{Robust Functional Projection},
abbreviated as RFP. For an analytic function $f$, we define
\begin{align}
\Delta_n(f)
=
\operatorname{tr}
\left\{
f(\bm C_n)(\bm I_p-\bm{\mathcal Z}_0)
\right\}
-
\frac{(p-r_n)^2}{2\pi\mathrm i}
\oint\nolimits_{\Gamma}
\frac{f(z)\,\mathrm dz}
{
z\Bigl[
p-r_n+
\underline m_n(z)
\operatorname{tr}
\left\{
\bm C_n(\bm I_p-\bm{\mathcal Z}_0)
\right\}
\Bigr]
},
\label{intro:fpplus}
\end{align}
where
\[
    \underline m_n(z)
    =
    \frac{1}{n}
    \operatorname{tr}
    \left(
      \frac{1}{n}\bm Y^{\prime}\bm Y-z\bm I_n
    \right)^{-1},
    \qquad
    \bm Y=(\bm y_1,\ldots,\bm y_n),
\]
and $\Gamma$ is a positively oriented contour enclosing the spectrum
of $\bm C_n$ within the analytic domain of $f$. Consequently, every quantity appearing
in \eqref{intro:fpplus} can be computed from the observations and the
given subspace.
Under $H_0$, when $p/n\to c\in(0,\infty)$ and $r_n/n\to0$, we establish
that
\begin{align}
\frac{
\Delta_n(f)-\widehat{\mu}_n(f)
}{
\widehat{\varrho}_n^{1/2}(f)
}
\ \xrightarrow{\mathrm d}
\mathcal N(0,1),
\label{intro:fpplusclt}
\end{align}
where $\widehat{\mu}_n(f)$ and $\widehat{\varrho}_n(f)$ are fully data-driven estimators of the asymptotic mean and variance. This result
provides a feasible eigenspace test without requiring knowledge of the radial distribution and the background level.

Alongside testing an entire spiked eigenspace, it is also of
interest to assess an individual candidate signal direction.
In particular, prior information may specify a single direction.
The question is then whether this candidate is an eigenvector
of the population scatter matrix associated with an eigenvalue
above the background level.
For a given unit vector $\bm\theta_0$, we therefore consider
\begin{align}
H_0^{\mathrm{RSD}}:\quad
&\bm T\bm\theta_0=\lambda\bm\theta_0
\quad\text{for some }\lambda>\sigma^2,
\notag\\
H_1^{\mathrm{RSD}}:\quad
&\bm T\bm\theta_0\ne\lambda\bm\theta_0
\quad\text{for all }\lambda>\sigma^2.
\label{intro:rsd}
\end{align}
Thus, the question is whether $\bm\theta_0$ is an eigenvector
associated with one of the spikes.
For fixed dimension $p$, \citet{paindaveine2020testing} study testing
for the leading principal direction under weak identifiability,
while \citet{paindaveine2020sign} and
\citet{paindaveine2026rank} investigate sign- and rank-based
tests, respectively, for this problem under elliptical
distributions. Our objective is to obtain feasible and robust
inference for a given spike direction in the high-dimensional
condition $p/n\to c\in(0,\infty)$, allowing the number of spikes
to grow and the spike increments to tend to zero.

We propose the \textbf{Robust Spike Direction} (RSD) test
to assess whether a given unit vector $\bm\theta_0$ is
an eigenvector of $\bm T$ associated with a spike eigenvalue.
Recalling the definitions of $\sigma^2$ and $d_{j,n}$ for
$j=1,\ldots,r_n$ in \eqref{intro:spiked-scatter}, we define
\[
\delta_n^{\prime}
=
\min_{1\le j\le r_n}\frac{d_{j,n}}{\sigma^2}.
\]
We allow $r_n=\mathrm o(p)$ spikes and assume that, for some
constant $K\ge1$ independent of $n$,
\begin{align}\label{www}
\delta_n^{\prime}
\le
\frac{d_{j,n}}{\sigma^2}
\le
K\delta_n^{\prime},
\qquad j=1,\ldots,r_n,
\end{align}
and
\begin{align}\label{qqq}
   n^{1/4} \delta_n^{\prime}\longrightarrow \infty
\end{align}
Thus, the relative spike increments ${d_{j,n}}/{\sigma^2}$ are of the same order.
The quantity $\delta_n$ describes the smallest relative spike increment between a spike eigenvalue and the background
level, and it may tend to zero.

The null hypothesis can fail in two ways:
$\bm\theta_0$ is not an eigenvector of $\bm T$, or it is
an eigenvector corresponding to the background eigenvalue
$\sigma^2$.
These two possibilities motivate complementary parts
of the RSD procedure.
To assess whether the given vector is an eigenvector of scatter matrix, we use
\begin{equation}
T_{Q,n}
=
\frac{
\sqrt n\left\{
\bm\theta_0^\prime\bm C_n^2\bm\theta_0
-
\left(\bm\theta_0^\prime\bm C_n\bm\theta_0\right)^2
-
\frac pn\bm\theta_0^\prime\bm C_n\bm\theta_0
\right\}
}{
\left(\bm\theta_0^\prime\bm C_n\bm\theta_0\right)
\sqrt{
\frac{2p}{n}
\left\{
\frac1p\operatorname{tr}(\bm C_n^2)-\frac pn
\right\}
}
}.
\label{eq:rsd-statistic-intro}
\end{equation}
The expression in braces in the numerator estimates
\[
\bm\theta_0^\prime\bm\Sigma^2\bm\theta_0
-\bigl(\bm\theta_0^\prime\bm\Sigma\bm\theta_0\bigr)^2
=
\bigl\|
(\bm I_p-\bm\theta_0\bm\theta_0^\prime)
\bm\Sigma\bm\theta_0
\bigr\|^2.
\]
This quantity is nonnegative and equals zero if and only if
$\bm\theta_0$ is an eigenvector
of $\bm T$.
The subtraction of
$(p/n)\bm\theta_0^\prime\bm C_n\bm\theta_0$
removes the leading bias caused by high-dimensional sampling so that the expression in braces in the numerator is asymptotically unbiased.
Under $H_0^{\mathrm{RSD}}$ and the conditions stated below, our paper proves
\[
T_{Q,n}\xrightarrow{d}\mathcal N(0,1).
\]

However, checking whether $\bm\theta_0$ is an eigenvector of $\bm T$
does not determine whether it corresponds to a spike or
background eigenvalue.
We therefore additionally compare
$\bm\theta_0^\prime\bm C_n\bm\theta_0$ with $1+s_n$,
where $s_n>0$ is a deterministic sequence satisfying
\begin{equation}
s_n=\mathrm o(\delta_n),
\qquad
\sqrt n\,\delta_n s_n\longrightarrow\infty.
\label{eq:rsd-threshold-intro}
\end{equation}
The RSD test rejects the null hypothesis $H_0^{\mathrm{RSD}}$  when
\begin{equation}
T_{Q,n}>z_{1-\alpha}
\quad\text{or}\quad
\bm\theta_0^\prime\bm C_n\bm\theta_0<1+s_n,
\label{eq:rsd-rejection-intro}
\end{equation}
where $z_{1-\alpha}$ denotes the $(1-\alpha)$-quantile
of the standard normal distribution.
The procedure is directly feasible.
Once $s_n$ is specified, all quantities are computed
from $\bm C_n$ and the candidate direction $\bm\theta_0$,
without estimating the number of spikes, their individual
eigenvalues, or the background level.

To establish the asymptotic distributions of both RFP and RSD,
 we develop a general theory of \textbf{G}eneralized \textbf{E}lliptical \textbf{S}patial-sign \textbf{S}pectral statistics (GESS), motivated by \citet{hu2024generalized}. Specifically, 
for an analytic function $f$ and a deterministic weighting matrix $\bm B_n$, we define the GESS by
\[
    L_n(f,\bm B_n)
    =
    \operatorname{tr}\{f(\bm C_n)\bm B_n\}.
\]
Here we allow for a more general matrix $\bm T$ than that in ]\eqref{intro:spiked-scatter}, requiring only its spectral norm to be bounded. The RFP statistic is covered by this class of statistics through
the choice $\bm B_n=\bm I_p-\bm{\mathcal Z}_0$, together with the
resolvent functions appearing in its contour integral. For RSD, we use the rank-one weighting matrix
$\bm B_n=\bm\theta_0\bm\theta_0^{\prime}$ and the functions
$f(z)=z$ and $f(z)=z^2$ to calculate the joint limit.

These constructions share a common idea: we construct statistics
for the hypothesized eigenspace or direction without first
seeking consistent estimators of the individual spike eigenvalues because
estimating individual spike eigenvalues in high dimensions
can require substantial restrictions. For example,
\citet{xie2026limiting} study the asymptotic behaviors of
sample eigenvalues corresponding to population spikes under elliptical distributions.
For bounded population spikes, they distinguish between
distant spikes and close spikes. Their central limit theorem
applies to sample eigenvalues corresponding to simple distant
spikes; these sample eigenvalues separate from the bulk.
For close spikes, the corresponding sample eigenvalues may
instead converge to a spectral edge or a quantile of the
limiting spectral distribution, so their limiting locations
do not generally identify individual population spikes.
Moreover, their results also require a concentration condition on the radial variable. This condition is stronger than requiring a finite fourth moment and is not satisfied by multivariate Student’s \(t\) distributions with fixed degrees of freedom.
These results illustrate the spectral and radial restrictions
that can arise when using individual sample eigenvalues
to estimate population spikes.

Our proposed two testing approaches avoid consistent estimation of individual spike
eigenvalues through different constructions. For RFP, we work with the hypothesized background subspace, on which the population spatial-sign covariance matrix has a common eigenvalue under $H_0$, thereby avoiding the need to estimate the individual spikes. For RSD, we directly construct a
statistic whose mean is asymptotically zero under $H_0^{\mathrm{RSD}}$,
allowing its null distribution to be constructed without
estimating the individual spike eigenvalues. We summarize our contributions as follows:
\begin{itemize}

\item We propose \textbf{Robust Functional Projection} (RFP)
for testing a given spiked eigenspace under high-dimensional
elliptical distributions. Like FP, RFP does not require
separation of the corresponding sample eigenvalues from the
bulk. It further allows the radial distribution and the
background level to be unknown.

\item We propose \textbf{Robust Spike Direction} (RSD) for
testing whether a given vector is a spike eigenvector.
The test provides robust directional inference when $p$
and $n$ are of the same order, allowing $r_n=\mathrm o(p)$ spikes
with comparable relative gaps from the background that may
tend to zero. The noise level and the remaining spike
directions need not be known.

\item We develop a general GESS theory for statistics
of the form $\operatorname{tr}\{f(\bm C_n)\bm B_n\}$,
providing a common foundation for the construction and
asymptotic analysis of RFP and RSD. We establish central
limit theorems covering both
$\operatorname{rank}(\bm B_n)\asymp p$ and
$\operatorname{rank}(\bm B_n)=\mathrm o(p)$ without imposing
moment conditions on the radial variable.

\item We apply RFP and RSD to LOCATA acoustic data.
RFP tests whether a candidate signal subspace constructed
from measured source and array geometry equals the population
spiked eigenspace, while RSD tests whether the corresponding
candidate direction is a spike eigenvector.

\end{itemize}

The remainder of the paper is organized as follows.
Section~\ref{sec:glss} establishes central limit theorems for
GESS with weighting matrices $\bm B_n$ of different rank orders,
providing the theoretical foundation for both testing procedures.
Section~\ref{sec:rfp} develops the RFP test  and
examines its finite-sample performance through simulations.
Section~\ref{sec:rsd} develops the RSD test for a given direction, and presents its asymptotic properties and
simulation results.
Finally, Section~\ref{sec:locata} applies both tests to LOCATA
acoustic data, using measured source and array geometry to
specify a candidate signal subspace and steering direction.
The two tests assess these geometric specifications at the
subspace and individual-direction levels, respectively.

\section{\textbf{Generalized elliptical spatial-sign spectral statistic}}\label{sec:glss}

The asymptotic analysis of RFP and RSD requires the fluctuations of spectral
statistics that retain information about a given subspace. This
motivates the following general theory for spatial-sign covariance
matrices under elliptical distributions. 

The asymptotic distribution of the GESS will be discussed when $\frac{k_n}{n}\rightarrow 0$ and $\frac{k_n}{n}\ge c_0$ for some positive constant $c_0$ respectively. Before delving into the main theorems in Section \ref{main results}, we provide an introduction to some preliminary results regarding the limiting spectral distribution of the conventional spatial-sign covariance
matrix $\bm C_n$ in Section \ref{preliminary}.
\subsection{Some preliminary results on the spatial-sign covariance matrix}\label{preliminary}
The Stieltjes transform is a fundamental function in random matrix theory, which is formally defined as 
	\begin{definition}\label{ST}
		For any function $G$ with bounded variation on the real line, its Stieltjes transform is defined by
		$$m_G(z)=\int \frac{1}{x-z}dG(x), \qquad z\in\mathbb{C} ~\text{and}~ \Im{z}\neq0.$$
	\end{definition}
It has been demonstrated that a bijective correspondence exists between $G$ and its Stieltjes transform $m_G(z)$ when $G$ is a proper distribution function (see Theorem B.8 in \citet{Bai:2010}). Recalling the definition of $\bm{C}_n$, an elementary limit theorem concerning the eigenvalues of $\bm{C}_n$ is focused on its empirical spectral distribution $F^{\bm{C}_n}$. To be more specific,
	$$
	F^{\bm{C}_n}(x)=\frac{1}{p}\sum_{i=1}^p\mathbb{I}_{\{\lambda_i(\bm{C}_n)\leq x\}}.
	$$
    If we assume that for all $n$, $\bm x_{i}$ are selected from elliptical distribution and $\bm y_i=\frac{\sqrt{p}\bm x_i}{\|\bm{x}_i\|}$, $F^{\bm{\Sigma}_p}$ convergences in distribution to $H$ and $p/n\rightarrow c\in (0,\infty)$, then almost surely, $F^{\bm{C}_n}$ converges in distribution to $F^{c, H}$, whose  Stieltjes transform $m(z)$ is the unique solution to
	\begin{equation}\label{m3}
		m(z)=\int \frac{1}{x(1-c-c z m(z))-z} d H(x), \quad z \in \mathbb{C}^{+}.
	\end{equation}
One can see more details of the proof of this assertion in \citet{li2017spectral}.
Considering $\bm{\underline{C}}_n \equiv(1 / n) \bm{Y}_n^{\prime}\bm{Y}_n$ whose spectrum differs from that of $\bm{C}_n$ by $|n-p|$ zeros, we know that its limiting empirical distribution function satisfies
	$$
	\underline{F}^{c, H} \equiv(1-c) \mathbb{I}_{[0, \infty)}+c F^{c, H}.
	$$
Furthermore, its Stieltjes transform
	\begin{equation}\label{m1}
		\underline{m}(z) \equiv m_{\underline{F}^{c, H}}(z)=-\frac{1-c}{z}+c m(z)
	\end{equation}
	has inverse
	\begin{equation}\label{m2}
		z=z(\underline{m})=-\frac{1}{\underline{m}}+c \int \frac{t}{1+t \underline{m}} d H(t),
	\end{equation}
	
	Let ${m}_n^0(z)$ and $\underline{m}_n^0(z)$ represent the quantities obtained from equations (\ref{m3}) and (\ref{m1}) when replacing $(c, H)$ by $(c_n, H_n)$, which will be frequently used in establishing our main theorems. The corresponding distribution functions for $m_n^0(z)$ and $\underline{m}_n^0(z)$ are denoted as $F^{c_n,H_n}$ and $\underline{F}^{c_n,H_n}$, respectively. In addition, $m_n(z)$ and $\underline{m}_n(z)$ are employed to denote the Stieltjes transforms of $F^{\bm{C}_n}$ and $F^{\underline{\bm{C}}_n}$. 
\subsection{Main theoretical results}\label{main results}
The following assumptions will be used in our theoretical analysis.
\begin{assumption} \label{asa}
    $n,p\to \infty$ and $\frac{p}{n}\to c$.
\end{assumption}
\begin{assumption}\label{asd}
    $(u_{j})_{j\geq1}$ are i.i.d variables selected from $U(S^{p-1})$
    and $\zeta _{j}$ are i.i.d scalars satisfying
    $$
    \mathbb{P}(\zeta=0)=0.
    $$
\end{assumption}
 \begin{assumption}\label{asb}
We set $\bm{\Sigma}_n=\mathbb{E}(\bm y\bm y^{\prime})$. The matrices $\bm{\Sigma}_n$, $\bm{T}_n$ and $\bm{B}_n$ are $p\times p$ non-random real symmetric matrices such that their non-zero eigenvalues are bounded away from $0$ and infinity and $\operatorname{tr}(\bm{\Sigma})=\mathrm{O}(p)$. Moreover, we assume that $\bm{\Sigma}_n$ is non-negative definite ($\bm{\Sigma}_n\succeq \bm{0}$) and  $H_n=F^{\bm{\Sigma}_n} \stackrel{D}{\rightarrow} H$, where $H$ is a distribution function.
 \end{assumption}
  \begin{assumption}\label{asc}
  Let $k_n=\operatorname{rank}(\bm{B}_n)$. Either one of the following two cases holds:

   (i) {[$k_n$ is comparable to $p$].} There exists a positive constant $c_0$ such that
		$\frac{k_n}{p}\geq c_0.$

  (ii) {[$k_n$ is much smaller than $p$].} $\frac{k_n}{p}\rightarrow  0.$
 \end{assumption}
 Assumptions \ref{asa} and \ref{asb} are standard in random matrix theory (see \citet{Bai:2007,Bai:2004,pan:2008} for example). The asymptotic behavior of GESS: $\operatorname{tr}f(\bm{C}_n)\bm{B}_n$ depends on the rank of $\bm{B}_n$. In the following Theorem \ref{main} and Theorem \ref{the6.1}, we summarize the different limiting distributions under the two different cases of $k_n$ stated in Assumption \ref{asc}, respectively. Assumption \ref{asd} is a general assumption on the elliptical distribution requiring no moment conditions on $\zeta$, enabling us to deal with heavy-tailed scenarios. In contrast, literature works such as \citet{hu2019high,ding2023extreme,jun2022tracy,yin2023central,xie2026limiting} impose some specific moment conditions and thus cannot handle statistical problems arising from extremely heavy-tailed distributions.

 \begin{theorem}\label{main} [$k_n$ is comparable to $n$]. Suppose that  Assumptions \ref{asa}, \ref{asd},\ref{asb} and \ref{asc} (i) hold.
	   Let $f_1, \ldots, f_r$ be analytic functions  on an open interval containing $[d_{-},d^{+}]$, where
		\begin{equation}\label{interval}
			[d_{-},d^{+}]=\left[\liminf_{n}\lambda_{\min}^{\bm{\Sigma}_n}\mathbb{I}_{(0,1)}(c)(1-\sqrt{c})^2, \limsup_{n}||\bm{\Sigma}_n||(1+\sqrt{c})^2\right].
		\end{equation}
Recall the definition of GESS defined previously, and define
\begin{equation}\label{add2-1}
  \Theta_n(f)=\operatorname{tr} f(\bm{C}_n)\bm{B}_n-\frac{1}{2\pi i}\oint_{\Gamma}f(z)\operatorname{tr}(z\bm{I}_p+z\underline{m}_n^0(z)\bm{\Sigma}_n)^{-1}\bm{B}_ndz,
\end{equation}
 where $\Gamma$ is a contour taken in the positive direction enclosing an open interval covering $[d_{-},d^{+}]$.
  Then we have the following results:
		
		\vskip 0.4cm
		(i) the random vector
		\begin{equation}\label{Vector}
			\left(\Theta_n(f_1), \ldots, \Theta_n(f_r)\right)
		\end{equation}
		forms a tight sequence in $n$.
		
		(ii)  After suitable centralization, the random vector (\ref{Vector}) converges weakly to an $r$-dimensional Gaussian distribution, i.e.,
		\begin{align}\label{converge}
			(\Theta_n(f_1)-\omega_{n,ell}(f_1),\cdots,\Theta_n(f_r)-\omega_{n,ell(f_r)})\stackrel{D}{\rightarrow}\mathcal{N}(\bm{0},\bm{\Omega}_{1,ell}).
		\end{align}
        We assume that the limits defining $\bm{\Omega}_{1,ell}$
exist and are finite, while explicit expressions for $\omega_{n,ell}(f)$ and $\bm{\Omega}_{1,ell}$ are given in  Supplementary Material.
		\end{theorem}

The asymptotic distribution of GESS is now investigated when $k_n=\mathrm{o}(p)$. It should be noted that when $k_n/n\rightarrow 0$, the quantities in $\bm{\Omega}_{1,ell}$ in Theorem \ref{main} all become zeros, resulting in $\Theta_n(f)=\mathrm o_P(1)$. Consequently, we need to find a suitable sequence $a_n \rightarrow\infty$, such that $a_n\Theta_n(f)$ converges to a non-degenerate distribution.
	\begin{theorem}\label{the6.1} [$k_n$ is much smaller than $n$].
Suppose that Assumptions \ref{asa}, \ref{asd},\ref{asb} and \ref{asc} (ii) hold, 
		then we have
		
		(i) the random vector
		\begin{equation}\label{Vector1}
			\sqrt{\frac{n}{k_n}}\left(\Theta_n(f_1), \ldots, \Theta_n(f_r)\right)
		\end{equation}
		forms a tight sequence in $n$.
		
		(ii) The random vector (\ref{Vector1}) converges weakly to a mean-zero $r$-dimensional Gaussian distribution, i.e.,
		\begin{align}\label{converge1}
				\sqrt{\frac n{k_n}}(\Theta_n(f_1),\cdots,\Theta_n(f_r))\stackrel{D}{\rightarrow}\mathcal{N}(\bm{0},\bm{\Omega}_2).
		\end{align}	
    Explicit formulas for the entries of $\bm{\Omega}_2$ are given in the supplement. These entries are assumed to have finite limits.
	\end{theorem}
    \begin{remark}
Some special cases of GESS have already been
studied. \citet{li2022eigenvalues} establish a central limit
theorem for LSS of spatial-sign covariance matrices,
corresponding to $\bm B_n=\bm I_p$.
\citet{xu2023eigenvectors} establish a central limit theorem
for eigenvector-weighted spectral statistics, which can be
written as $\bm v_n^{\prime}f(\bm C_n)\bm v_n$ for a
deterministic unit vector $\bm v_n$, corresponding to
$\bm B_n=\bm v_n\bm v_n^{\prime}$.
Our results consider more general deterministic weighting
matrices under the elliptical model, with ranks either
comparable to the dimension or of smaller order.
\end{remark}
\section{\textbf{Robust functional projection}}\label{sec:rfp}

We use the GESS theory developed above to test whether
the population spiked eigenspace equals a given subspace.
Under the elliptical model, we suppose that
\begin{equation}
\bm T_n
=
\sigma_n^2\bm I_p
+
\bm V_n\bm D_n\bm V_n^{\prime},
\qquad
\bm V_n^{\prime}\bm V_n=\bm I_{r_n},
\qquad
\bm D_n=\operatorname{diag}(d_{1,n},\ldots,d_{r_n,n}),
\label{rfp:model}
\end{equation}
where $\sigma_n^2>0$ is the unknown background level and
$d_{j,n}>0$ for $j=1,\ldots,r_n$.
The columns of $\bm V_n$ span the spiked eigenspace,
whose orthogonal projection matrix is
$\mathcal Z_n=\bm V_n\bm V_n^{\prime}$.
We allow the number of spikes $r_n$ to grow with $r_n=\mathrm o(p)$.
For a given orthogonal projection matrix $\mathcal Z_0$, we consider
\begin{equation}
H_0:\mathcal Z_n=\mathcal Z_0,
\qquad\text{versus}\qquad
H_1:\mathcal Z_n\ne\mathcal Z_0,
\label{rfp:hypotheses}
\end{equation}
in the high-dimensional setting (i.e $p/n\rightarrow c\in (0,\infty)$) under elliptical distributions.
Under $H_0$, the orthogonal complement of the given subspace
contains only background directions. This observation motivates
a test based on the projection of $f(\bm C_n)$ onto that
complement, leading to the GESS
$\operatorname{tr}\{f(\bm C_n)(\bm I_p-\mathcal Z_0)\}$.

\subsection{Methodology}

We first establish that spatial normalization preserves
the spiked structure of the population matrix.

\begin{proposition}
\label{pro1}
Under the elliptical model, suppose that $\bm T_n$
admits the decomposition \eqref{rfp:model}.
Then the population spatial-sign covariance matrix
$\bm\Sigma_n$ admits the decomposition
\begin{equation}
\bm\Sigma_n
=
\tau_n\bm I_p
+
\bm V_n\widetilde{\bm D}_n\bm V_n^{\prime},
\qquad
\widetilde{\bm D}_n
=
\operatorname{diag}
(\widetilde d_{1,n},\ldots,\widetilde d_{r_n,n}),
\label{rfp:sscm-model}
\end{equation}
where $\tau_n>0$ and
$\widetilde d_{j,n}>0$ for $j=1,\ldots,r_n$.
\end{proposition}

Proposition~\ref{pro1} ensures that spatial normalization
preserves both the spiked eigenspace and the common
background eigenvalue structure, although it changes
the eigenvalues themselves. Consequently, under $H_0$,
\begin{equation}
\bm\Sigma_n(\bm I_p-\mathcal Z_0)
=
\tau_n(\bm I_p-\mathcal Z_0).
\label{rfp:null-background}
\end{equation}
The unknown background level can therefore be estimated
from the observations projected onto the orthogonal
complement of the given subspace.  We define
\begin{equation}
\widehat\tau_n
=
\frac{
\operatorname{tr}\{\bm C_n(\bm I_p-\mathcal Z_0)\}
}{p-r_n}
=
\frac{1}{n(p-r_n)}
\sum_{i=1}^{n}
\|(\bm I_p-\mathcal Z_0)\bm y_i\|^2.
\label{rfp:background-estimator}
\end{equation}
This estimator is unbiased for $\tau_n$ under $H_0$. Since $\|\bm y_i\|^2=p$,
\begin{equation}
\widehat\tau_n
=
\frac{
p-n^{-1}\sum_{i=1}^{n}\|\mathcal Z_0\bm y_i\|^2
}{p-r_n}.
\label{rfp:background-identity}
\end{equation}
Thus its fluctuations are determined by the projection
onto the hypothesized spiked space. The following result
establishes the precision needed for constructing RFP.

\begin{proposition}
\label{rfp:background-rate}
Under $H_0$ and the spectral boundedness assumptions
of the preceding section, suppose that
$p/n\to c\in(0,\infty)$ and $r_n=\mathrm o(p)$. Then
\begin{equation}
\widehat\tau_n-\tau_n
=
\mathrm{o}_{P}(p^{-1}).
\label{rfp:background-precision}
\end{equation}
\end{proposition}

We define the robust functional projection statistic by
\begin{equation}
\begin{split}
\Delta_n(f)
={}&
\operatorname{tr}
\{f(\bm C_n)(\bm I_p-\mathcal Z_0)\}
-
\frac{p-r_n}{2\pi\mathrm{i}}
\oint_{\Gamma}
\frac{f(z)}
     {z\{1+\widehat\tau_n\underline m_n(z)\}}
\,\mathrm{d}z,
\end{split}
\label{rfp:statistic}
\end{equation}
where $\Gamma$ is a positively oriented contour chosen
as in the preceding GESS results.

Proposition~\ref{rfp:background-rate} provides the precision
needed to estimate the unknown background level of the spatial-sign covariance matrix. The background correction in \eqref{rfp:statistic}
aggregates the contributions of $p-r_n$ background directions
and therefore carries a factor $p-r_n\asymp p$.
 Replacing $\tau_n$ by
$\widehat\tau_n$ introduces an error of order
$\mathrm{O}_{ P}\{p|\widehat\tau_n-\tau_n|\}$.
Consequently, ordinary consistency,
$\widehat\tau_n-\tau_n=\mathrm{o}_{ P}(1)$,
does not by itself ensure that estimating the background
has a negligible effect on the test.
Proposition~\ref{rfp:background-rate} gives the stronger
conclusion
$(p-r_n)(\widehat\tau_n-\tau_n)=\mathrm{o}_{ P}(1)$.
The resulting statistic thus differs from its counterpart
using the true $\tau_n$ by $\mathrm{o}_{ P}(1)$.
 Thus, replacing $\tau_n$ by its estimator does not change
the asymptotic null distribution of the test statistic.

The first term in (\ref{rfp:statistic}) measures the spectral contribution in
directions orthogonal to the given subspace. The second term estimates the centering term under $H_0$
using $\underline m_n(z)$ and $\widehat\tau_n$.
Subtracting this estimate from the first term gives
the statistic $\Delta_n(f)$.
Both terms are computed from
$\bm C_n$ and $\mathcal Z_0$, with the unknown background
level estimated through \eqref{rfp:background-estimator}.

We next establish the asymptotic null distribution of
$\Delta_n(f)$. The empirical transform $\underline m_n(z)$
in \eqref{rfp:statistic} contributes to the fluctuations
of the statistic and must be analyzed jointly with the
projected spectral term. The GESS theory provides
the required limit results.

Let $\mu_n(f)$ and $\varrho_n(f)$ denote the asymptotic
mean and variance of $\Delta_n(f)$, respectively.
Their explicit contour-integral representations are
given in the Supplementary Material.

\begin{theorem}
\label{rfp:null-clt}
Suppose that the assumptions of the high-rank
GESS theorem hold and that
\eqref{rfp:model} is satisfied with $r_n=o(p)$.
Let \(f\) be analytic on a complex neighborhood of the relevant spectral interval and real-valued on that interval, and suppose that
$\liminf_{n\to\infty}\varrho_n(f)>0$.
Then, under $H_0$,
\begin{equation}
\frac{\Delta_n(f)-\mu_n(f)}
     {\sqrt{\varrho_n(f)}}
\xrightarrow{\mathcal D}
\mathcal N(0,1).
\label{rfp:oracle-clt}
\end{equation}
\end{theorem}

To implement the test, the unknown quantities appearing
in $\mu_n(f)$ and $\varrho_n(f)$ must also be estimated.
In addition to the background estimator
$\widehat\tau_n$, we use the trace normalization
$\operatorname{tr}(\bm\Sigma_n)=p$, which gives
\begin{equation}
\frac{1}{r_n}
\sum_{j=1}^{r_n}\widetilde d_{j,n}
=
\frac{p}{r_n}(1-\tau_n).
\label{rfp:average-spike}
\end{equation}
This motivates the estimator
\begin{equation}
\widehat d_n
=
\frac{p}{r_n}(1-\widehat\tau_n)
\label{rfp:spike-estimator}
\end{equation}
of the average spike increment.

We obtain $\widehat\mu_n(f)$ and
$\widehat\varrho_n(f)$ by making the substitutions
\begin{equation}
\tau_n\mapsto\widehat\tau_n,
\qquad
\widetilde d_{j,n}\mapsto\widehat d_n
\quad (j=1,\ldots,r_n),
\qquad
\underline m_n^0(z)\mapsto\underline m_n(z)
\label{rfp:substitutions}
\end{equation}
in the expressions for $\mu_n(f)$ and $\varrho_n(f)$.
Here $\underline m_n^0(z)$ is defined in the preceding section. Traces in these expressions are estimated by replacing the background eigenvalue with $\widehat\tau_n$ and each spike eigenvalue with $\widehat\tau_n+\widehat d_n$.

The common replacement $\widehat d_n$ does not require
the spike increments to be equal, nor does it estimate
each increment consistently. Its use is justified by
the structure of the asymptotic mean and variance:
the individual spike increments enter through
normalized sums whose contributions are of order
$r_n/p$. Under the boundedness assumptions and
$r_n=\mathrm o(p)$, replacing these increments by their
estimated average produces a negligible error.
This is the same assertion used in the functional
projection approach of \citet{hu2024generalized}.
The additional replacement of the unknown background
level is justified by
Proposition~\ref{rfp:background-rate}.

Under the conditions of Theorem~\ref{rfp:null-clt}, we obtain
\begin{equation}
\widehat\mu_n(f)-\mu_n(f)
=
\mathrm{o}_{ P}(1),
\qquad
\widehat\varrho_n(f)-\varrho_n(f)
=
\mathrm{o}_{ P}(1).
\label{rfp:parameter-consistency}
\end{equation}
Consequently, the standardized RFP statistic
\begin{equation}
T_{\mathrm{RFP},n}(f)
=
\frac{\Delta_n(f)-\widehat\mu_n(f)}
     {\sqrt{\widehat\varrho_n(f)}}
\label{rfp:standardized-statistic}
\end{equation}
satisfies
\begin{equation}
T_{\mathrm{RFP},n}(f)
\xrightarrow{\mathcal D}
\mathcal N(0,1)
\qquad\text{under }H_0.
\label{rfp:feasible-clt}
\end{equation}
For a fixed choice of $f$, a two-sided test with
asymptotic significance level $\alpha$ therefore
rejects $H_0$ when
\begin{equation}
\bigl|T_{\mathrm{RFP},n}(f)\bigr|
>
z_{1-\alpha/2}.
\label{rfp:rejection-rule}
\end{equation}
 The resulting procedure allows
the background level to be unknown and requires no
moment conditions on the radial variable.

\subsection{Simulation}
\label{subsec:rfp-simulation}

We examine the finite-sample size and power of RFP with
$n=1000$, $p\in\{500,1000\}$ and a nominal significance
level of $0.05$.
Let $\bm e_j$ denote the $j$th coordinate vector in $\mathbb R^p$.
The given subspace has orthonormal basis
$\bm V_0=(\bm e_1,\ldots,\bm e_{r_n})$ and projection matrix
$\bm Z_0=\bm V_0\bm V_0^{\prime}$.
Under the null hypothesis, we take the scatter matrix
\begin{equation}
\bm T_0
=
\operatorname{diag}
\bigl(
1+d_1,\ldots,1+d_{r_n},
\underbrace{1,\ldots,1}_{p-r_n}
\bigr),
\label{eq:sim-scatter}
\end{equation}
where $d_1,\ldots,d_{r_n}>0$ are the spike increments above
the unit background.
We generate independent observations from a multivariate
Cauchy distribution:
\begin{equation}
\bm x_i
=
\frac{\bm T_0^{1/2}\bm g_i}{\sqrt{w_i}},
\qquad
\bm g_i\sim N_p(\bm 0,\bm I_p),
\qquad
w_i\sim\chi^2_1,
\label{eq:sim-elliptical-data}
\end{equation}
where $\bm g_i$ and $w_i$ are independent.
Although the population covariance matrix does not exist,
RFP still permits inference on the spiked eigenspace of the
scatter matrix. For each simulated sample, we estimate the unknown background
eigenvalue $\tau_n$ from the data and use this estimate to
compute the standardized RFP statistic.
We perform a two-sided test at significance level $\alpha=0.05$,
rejecting $H_0$ when the absolute value of the standardized
statistic exceeds $z_{1-\alpha/2}$.

\begin{figure}[!htbp]
\centering
\includegraphics[width=0.75\linewidth]{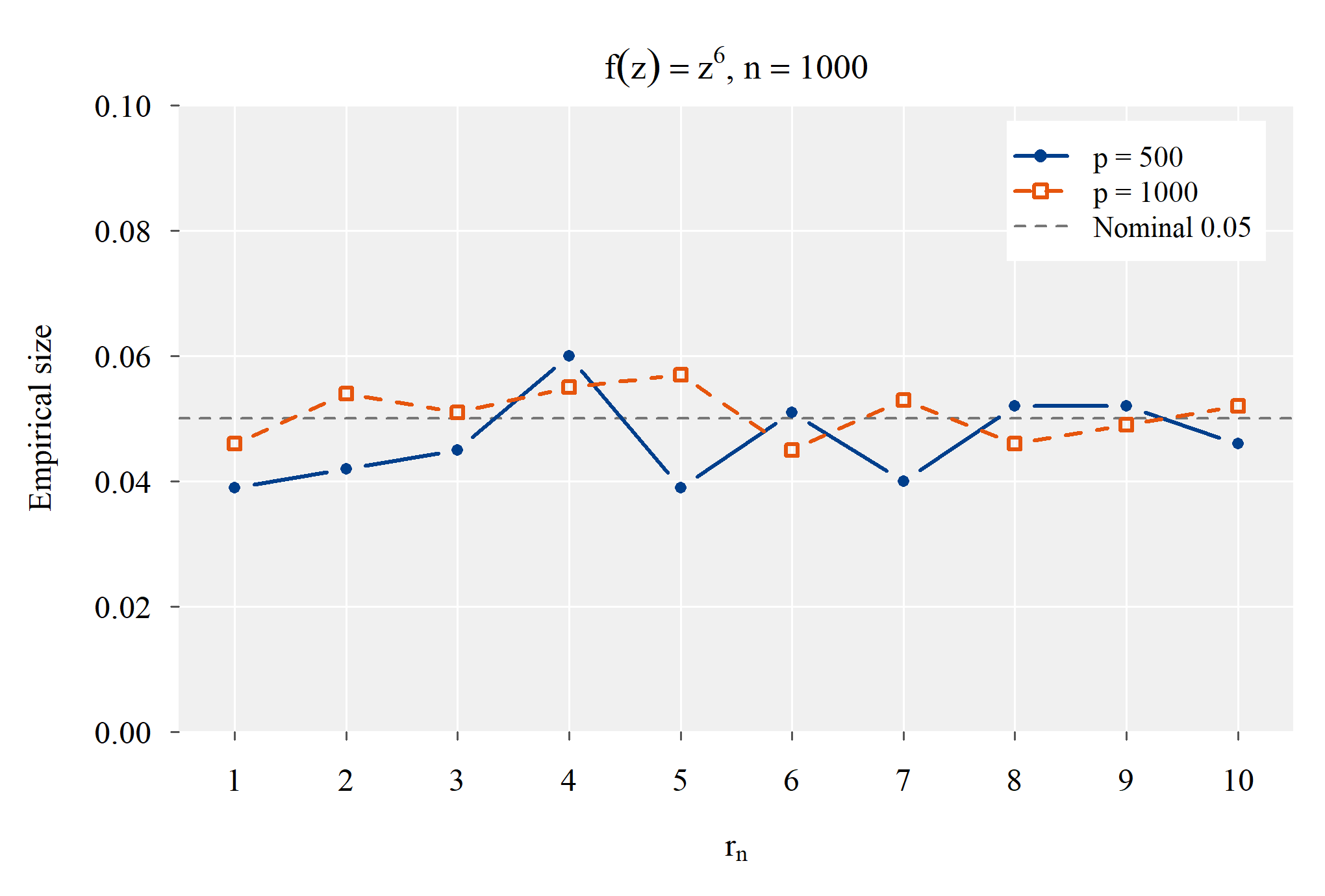}
\caption{Empirical size of RFP with $f(z)=z^6$,
$n=1000$ and $p\in\{500,1000\}$ as the number of spikes varies.
All spike increments equal $1$, and each experiment uses
$1000$ replications. The dashed horizontal line indicates
the nominal level $0.05$.}
\label{fig:rfp-size}
\end{figure}

We first assess size as the number of spikes varies.
We set $d_1=\cdots=d_{r_n}=1$ and let $r_n$ range from $1$ to $10$.
Each configuration uses $1000$ replications.
Figure~\ref{fig:rfp-size} shows the empirical rejection
rates for $f(z)=z^6$.
For both dimensions, the empirical sizes remain close to
the nominal level over the range of ranks considered.

\FloatBarrier

To study power, we rotate the leading spike direction towards
a background direction while keeping the remaining spike
directions fixed. Specifically, let
\begin{equation}
\bm v_1(\theta)
=
\cos(\theta)\bm e_1+\sin(\theta)\bm e_{r_n+1},
\qquad
\bm V_\theta
=
\bigl(\bm v_1(\theta),\bm e_2,\ldots,\bm e_{r_n}\bigr),
\label{eq:sim-rotation}
\end{equation}
and replace $\bm T_0$ in \eqref{eq:sim-elliptical-data} by
\begin{equation}
\bm T_\theta
=
\bm I_p+
\bm V_\theta
\operatorname{diag}(d_1,\ldots,d_{r_n})
\bm V_\theta^{\prime}.
\label{eq:sim-alternative-scatter}
\end{equation}
The given projection matrix remains $\bm Z_0$.
Thus, $\theta=0$ corresponds to the null hypothesis,
whereas a nonzero angle in the range considered below changes
the population spiked eigenspace.

We consider two configurations:
\begin{align}
\text{Setting 1:}\quad
& r_n=3,
\qquad (d_1,d_2,d_3)=(6,5.5,5),
\notag\\
\text{Setting 2:}\quad
& r_n=8,
\qquad (d_1,\ldots,d_8)=(9,1,1,1,1,1,1,1).
\label{eq:sim-power-settings}
\end{align}
For each setting and dimension, we evaluate $f(z)=z^5$ and
$f(z)=z^6$ on the same simulated samples at
$\theta\in\{0^\circ,1^\circ,\ldots,15^\circ,30^\circ\}$.
We use $1000$ replications at $\theta=0$ and $200$
replications at each nonzero angle.

Both functions perform well in the two simulation settings.
The empirical power increases rapidly with the rotation angle
across all experiments considered. At $\theta=15^\circ$,
both functions achieve empirical power of at least $0.99$
in every experiment.

\begin{figure}[H]
\centering
\includegraphics[width=\linewidth]{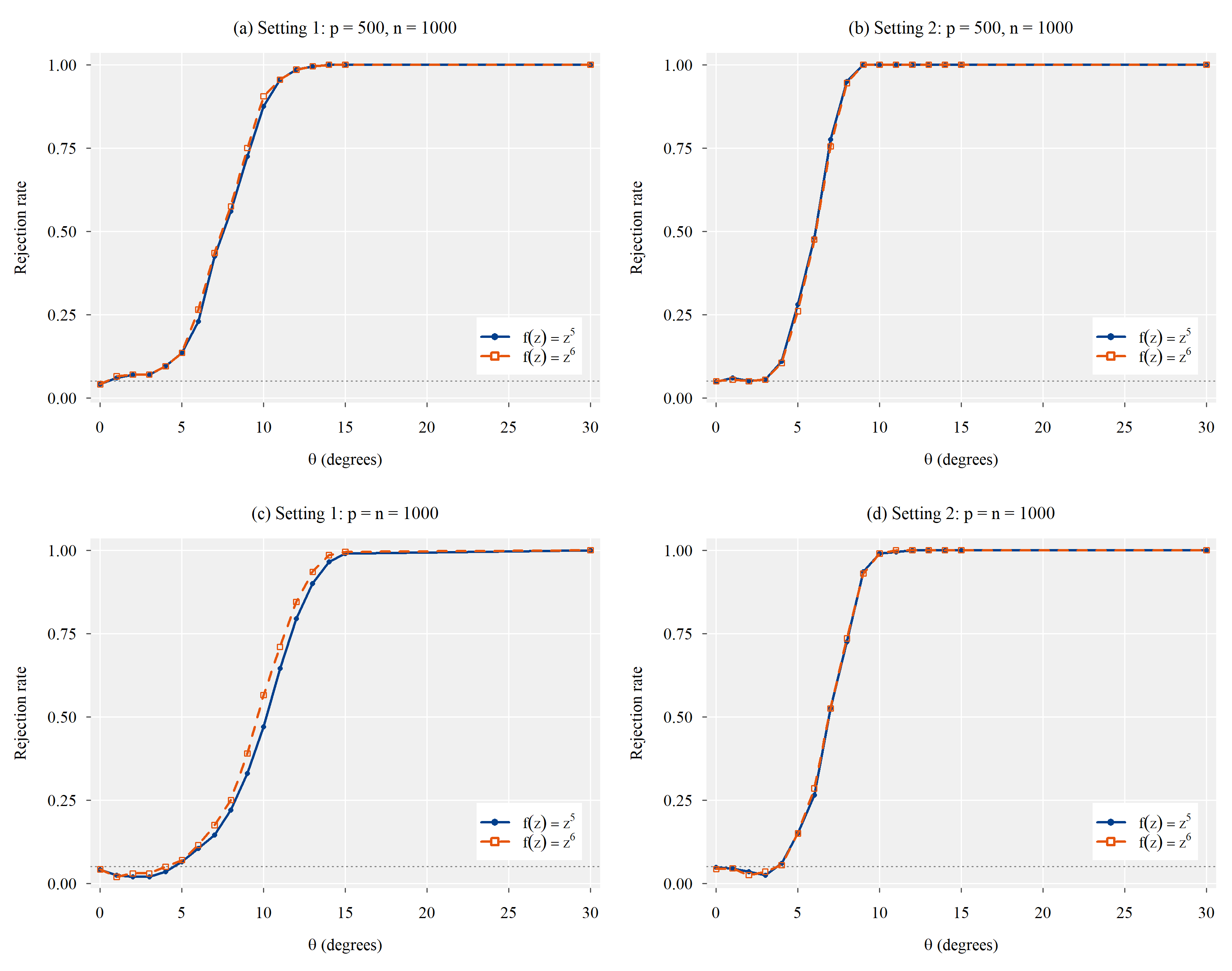}
\caption{Empirical rejection frequencies of RFP with
$f(z)=z^5$ and $f(z)=z^6$, with $n=1000$ throughout.
}
\label{fig:rfp-power}
\end{figure}
\FloatBarrier
The left and right columns correspond to Settings~1 and~2,
respectively; the top and bottom rows correspond to
$p=500$ and $p=1000$, respectively.
The dashed horizontal lines indicate the nominal level $0.05$.

\section{\textbf{Robust spike direction}}
\label{sec:rsd}
We now develop RSD for testing whether a given unit vector is
a spike eigenvector under the elliptical spiked model.
Its construction and asymptotic theory rely on the
GESS results established above.

\subsection{Methodology}
\label{sec:rsd-methodology}

Let $\bm\theta_0\in\mathbb R^p$ be a given unit vector.
We consider
\begin{align}
H_0 &: \bm\Sigma_n\bm\theta_0=\lambda\bm\theta_0
\quad\text{for some }\lambda>\tau_n,
\notag\\
H_1 &: \bm\Sigma_n\bm\theta_0\ne\lambda\bm\theta_0
\quad\text{for every }\lambda>\tau_n,
\label{rsd:hypotheses}
\end{align}
where $\tau_n$ denotes the background eigenvalue.
Thus, the null hypothesis requires $\bm\theta_0$ itself to be
a spike eigenvector, while the alternative has two possibilities:
$\bm\theta_0$ is not an eigenvector of $\bm\Sigma_n$,
or it is an eigenvector corresponding to the background
eigenvalue. The RSD procedure addresses both possibilities.

The construction starts from the quantity
\begin{align}
q_n(\bm\theta_0)
&=
\bm\theta_0^\prime\bm\Sigma_n^2\bm\theta_0
-\bigl(\bm\theta_0^\prime\bm\Sigma_n\bm\theta_0\bigr)^2
=
\left\|
(\bm I_p-\bm\theta_0\bm\theta_0^\prime)
\bm\Sigma_n\bm\theta_0
\right\|^2.
\label{rsd:population-discrepancy}
\end{align}
This quantity is zero when $\bm\theta_0$ is an eigenvector
of $\bm\Sigma_n$ and positive otherwise. For example, a
combination of two spike eigenvectors with different
eigenvalues gives a positive value when both coefficients
are nonzero. However, the quantity also vanishes for
background eigenvectors, so it cannot by itself determine
whether $\bm\theta_0$ is a spike eigenvector.

To estimate $q_n(\bm\theta_0)$, we remove the leading
high-dimensional bias by subtracting
$c_n\bm\theta_0^\prime\bm C_n\bm\theta_0$ from
$\bm\theta_0^\prime\bm C_n^2\bm\theta_0
-(\bm\theta_0^\prime\bm C_n\bm\theta_0)^2$,
where $c_n=p/n$.
Under the spectral assumptions below, the remaining
centering error is of order $n^{-1}$ and is negligible
on the $n^{-1/2}$ fluctuation scale.
Moreover, we define
\begin{equation}
\widehat\gamma_{2,n}
=\frac1p\operatorname{tr}(\bm C_n^2)-c_n,
\qquad
\gamma_{2,n}
=\frac1p\operatorname{tr}(\bm\Sigma_n^2).
\label{rsd:gamma-estimator}
\end{equation}
We construct the statistic
\begin{equation}
T_{Q,n}
=
\frac{
\sqrt n\left\{
\bm\theta_0^\prime\bm C_n^2\bm\theta_0
-\bigl(\bm\theta_0^\prime\bm C_n\bm\theta_0\bigr)^2
-c_n\bm\theta_0^\prime\bm C_n\bm\theta_0
\right\}
}{
\bigl(\bm\theta_0^\prime\bm C_n\bm\theta_0\bigr)
\sqrt{2c_n\widehat\gamma_{2,n}}
}.
\label{rsd:quadratic-component}
\end{equation}
The normalization follows from the joint GESS
limit for $f(z)=z$ and $f(z)=z^2$, with weighting matrix
$\bm\theta_0\bm\theta_0^\prime$.
Large positive values of $T_{Q,n}$ provide evidence
against the assertion that $\bm \theta_0$ is an eigenvector of $\bm \Sigma_n$.

To distinguish background directions from spike eigenvectors,
we also examine the quadratic form
\[
a_n=\bm\theta_0^\prime\bm\Sigma_n\bm\theta_0.
\]
We denote the spike eigenvalues as
$\lambda_{j,n}=\tau_n+\widetilde d_{j,n}$,
$j=1,\ldots,r_n$, where $\widetilde d_{j,n}>0$, and define
\begin{equation}
\delta_n=\min_{1\le j\le r_n}\widetilde d_{j,n}.
\label{rsd:spike-strength}
\end{equation}
Thus, $\delta_n$ is the smallest gap between a spike
eigenvalue and the background eigenvalue of $\bm\Sigma_n$.

The signal condition considered in this paper is
\begin{equation}
\max_{1\le j\le r_n}\widetilde d_{j,n}\le K\delta_n,
\qquad n^{1/4}\delta_n\longrightarrow\infty,
\label{rsd:signal-condition}
\end{equation}
for some constant $K\ge1$ independent of $n$.
This condition requires the spike increments to have
the same order and to dominate $n^{-1/4}$.
It permits vanishing spikes and imposes no separation
between the spike eigenvalues themselves.
In particular, repeated spike eigenvalues are allowed.

\begin{proposition}\label{pro:rsd-spike-equivalence}
 Let $\sigma^2+d_{j,n}$ and
$\tau_n+\widetilde d_{j,n}$, $j=1,\ldots,r_n$, denote the spike
eigenvalues of $\bm T_n$ and its population spatial-sign covariance
matrix $\bm\Sigma_n$, respectively, where $\sigma^2$ and $\tau_n$
are the corresponding background eigenvalues.
Then \eqref{rsd:signal-condition} holds if and only if, for some
constant $K'\ge1$ independent of $n$,
\begin{equation}\label{rsd:equivalent-strength-condition}
    \max_{1\le j\le r_n}\frac{d_{j,n}}{\sigma^2}
    \le K'\min_{1\le j\le r_n}\frac{d_{j,n}}{\sigma^2},
    \qquad
    n^{1/4}\min_{1\le j\le r_n}\frac{d_{j,n}}{\sigma^2}
    \longrightarrow\infty.
\end{equation}
Moreover, under either condition,
\begin{equation}\label{rsd:spike-equivalence}
    \max_{1\le j\le r_n}
    \left|
    \frac{\widetilde d_{j,n}}{d_{j,n}/\sigma^2}-1
    \right|
    \longrightarrow0.
\end{equation}
\end{proposition}
This proposition shows that the condition \eqref{rsd:signal-condition}
is equivalent to the one stated in \eqref{www} and \eqref{qqq}
in Section \ref{Introductio}.
We continue to consider
$\bm\theta_0'\bm\Sigma_n\bm\theta_0$, estimated by
$\bm\theta_0'\bm C_n\bm\theta_0$. This quantity equals
$\tau_n$ for a background eigenvector and is at least
$\tau_n+\delta_n$ for a spike eigenvector, where
$\delta_n$ is defined in (\ref{rsd:spike-strength}).
 Since $\operatorname{tr}(\bm\Sigma_n)=p$, we have
\begin{equation}
\tau_n
=1-\frac1p\sum_{j=1}^{r_n}\widetilde d_{j,n}\le1.
\label{rsd:background-level}
\end{equation}
A background direction therefore has $a_n=\tau_n\le1$.
Under $r_n=\mathrm{o}(p)$ and the comparability condition
in \eqref{rsd:signal-condition},
\[
\min_{1\le j\le r_n}\lambda_{j,n}-1
=\delta_n-\frac1p\sum_{j=1}^{r_n}\widetilde d_{j,n}
=\delta_n\{1+\mathrm{o}(1)\}.
\]
This permits a threshold between the background and
all spike eigenvalues without estimating their
individual values, and
we choose a deterministic sequence $s_n>0$ satisfying
\begin{equation}
s_n=\mathrm{o}(\delta_n),
\qquad \sqrt n\,\delta_n s_n\longrightarrow\infty.
\label{rsd:threshold-rate}
\end{equation}
The first condition ensures that, for all sufficiently
large $n$, the threshold $1+s_n$ lies above the background
eigenvalue and below every spike eigenvalue.
We therefore also reject $H_0$ whenever
\begin{equation}
\bm\theta_0^\prime\bm C_n\bm\theta_0<1+s_n.
\label{rsd:background-component}
\end{equation}
The additional rejection condition in
(\ref{rsd:background-component}) provides power against
background directions and can also contribute to rejection
for directions close to the background eigenspace.
The second condition in \eqref{rsd:threshold-rate}
ensures that $T_{Q,n}\to+\infty$ in probability for
alternatives satisfying
\[
\bm\theta_0^\prime\bm\Sigma_n\bm\theta_0
=1+s_n+\mathrm O(n^{-1/2}).
\]
For these alternatives,
$\bm\theta_0'\bm\Sigma_n\bm\theta_0$ differs from the
threshold $1+s_n$ by $\mathrm O(n^{-1/2})$. Meanwhile,
\[
\bm\theta_0'\bm C_n\bm\theta_0
-\bm\theta_0'\bm\Sigma_n\bm\theta_0
=\mathrm O_P(n^{-1/2}).
\]
Thus, the estimation error is not negligible relative to
the distance from the threshold. We therefore cannot
guarantee that
$\bm\theta_0'\bm C_n\bm\theta_0<1+s_n$
holds with probability tending to one.
For these alternatives, the second condition in
(\ref{rsd:threshold-rate}) ensures that
$T_{Q,n}\to+\infty$ in probability, so the RSD test
still rejects with probability tending to one.

The admissible choices of $s_n$ depend on the
order of $\delta_n$.
For spikes of constant order, $\delta_n\asymp1$,
any choice $s_n=n^{-a}$ with $0<a<1/2$ satisfies
\eqref{rsd:threshold-rate}; examples include
$n^{-1/4}$ and $n^{-1/6}$.
If instead $\delta_n\asymp n^{-\beta}$ with
$0<\beta<1/4$, a choice $s_n=n^{-a}$ is admissible whenever
\[
\beta<a<\frac12-\beta.
\]
In particular, $s_n=n^{-1/4}$ satisfies
\eqref{rsd:threshold-rate} throughout the condition
$n^{1/4}\delta_n\to\infty$.
Thus, this choice accommodates vanishing spikes,
while other rates are available when the spikes are
known to be stronger.

For a nominal significance level $\alpha\in(0,1)$,
the RSD test rejects $H_0$ when
\begin{equation}
T_{Q,n}>z_{1-\alpha}
\quad\text{or}\quad
\bm\theta_0^\prime\bm C_n\bm\theta_0<1+s_n.
\label{rsd:rejection-rule}
\end{equation}
If the observed quadratic form is at least $1+s_n$,
the decision is determined solely by $T_{Q,n}$.
For a specified $s_n$, the rule uses only $\bm C_n$
and the given direction $\bm\theta_0$, without knowledge
of  the background level, or the
individual spike eigenvalues.

\begin{theorem}
\label{thm:rsd-null}
Suppose that Assumptions~\ref{asa}, \ref{asd} and \ref{asb}
hold, with $p/n\to c\in(0,\infty)$ and
$r_n=\mathrm{o}(p)$.
Assume the signal condition \eqref{rsd:signal-condition},
and let $s_n>0$ satisfy \eqref{rsd:threshold-rate}.
Then, under $H_0$,
\begin{equation}
T_{Q,n}\xrightarrow{d}\mathcal N(0,1),
\qquad
\mathbb P\!\left\{
\bm\theta_0^\prime\bm C_n\bm\theta_0<1+s_n
\right\}\longrightarrow0.
\label{rsd:null-limits}
\end{equation}
Consequently, the rejection rule \eqref{rsd:rejection-rule}
has asymptotic level $\alpha$:
\begin{equation}
\mathbb P\!\left\{
T_{Q,n}>z_{1-\alpha}
\ \text{or}\
\bm\theta_0^\prime\bm C_n\bm\theta_0<1+s_n
\right\}\longrightarrow\alpha.
\label{rsd:null-rejection-probability}
\end{equation}
\end{theorem}

Since the additional rejection event has probability $\mathrm{o}(1)$
under $H_0$, no adjustment to the one-sided critical value
$z_{1-\alpha}$ is needed.

Although the estimator of $q_n(\bm\theta_0)$ used above is
asymptotically unbiased, it may still be biased in finite samples.
We therefore apply a finite-sample correction to obtain an exactly
unbiased estimator and estimate its variance under $H_0$
using the spatial signs.
Define
\begin{equation}
\widehat q_{U,n}
=\frac{n}{n-1}\left\{
\bm\theta_0^\prime\bm C_n^2\bm\theta_0
-\bigl(\bm\theta_0^\prime\bm C_n\bm\theta_0\bigr)^2
-c_n\bm\theta_0^\prime\bm C_n\bm\theta_0
+\frac{1}{n^2}\sum_{i=1}^n
 (\bm\theta_0^\prime\bm y_i)^4
\right\},
\label{rsd:unbiased-discrepancy}
\end{equation}
where $\bm y_i$ are the spatial signs used to form
$\bm C_n$.
This estimator is exactly unbiased for
$q_n(\bm\theta_0)$.
Let
\begin{equation}
\bm z_i
=(\bm\theta_0^\prime\bm y_i)
(\bm I_p-\bm\theta_0\bm\theta_0^\prime)\bm y_i,
\qquad
\widehat\eta_n
=\frac{1}{n(n-1)}\sum_{i\ne j}
(\bm z_i^\prime\bm z_j)^2.
\label{rsd:null-variance-estimator}
\end{equation}
Under $H_0$, $2\widehat\eta_n/\{n(n-1)\}$ is an unbiased
estimator of $\operatorname{Var}(\widehat q_{U,n})$.
The corrected statistic is
\begin{equation}
T_{Q,n}^{\mathrm{corr}}
=\frac{\widehat q_{U,n}}
 {\{2\widehat\eta_n/[n(n-1)]\}^{1/2}}.
\label{rsd:corrected-statistic}
\end{equation}
The following proposition establishes its asymptotic
equivalence to $T_{Q,n}$.

\begin{proposition}
\label{pro:rsd-correction}
Suppose that the assumptions of Theorem~\ref{thm:rsd-null} hold.
For any deterministic sequence of unit vectors $\bm\theta_0$,
\begin{equation}
T_{Q,n}^{\mathrm{corr}}
=\{1+\mathrm{o}_P(1)\}T_{Q,n}+\mathrm{o}_P(1).
\label{rsd:correction-equivalence}
\end{equation}
This relation holds under both the null hypothesis and
the alternative sequences considered in
Theorem~\ref{thm:rsd-power}.
In particular, under $H_0$,
\begin{equation}
T_{Q,n}^{\mathrm{corr}}-T_{Q,n}\xrightarrow{P}0,
\qquad
T_{Q,n}^{\mathrm{corr}}\xrightarrow{d}\mathcal N(0,1).
\end{equation}
Replacing $T_{Q,n}$ by $T_{Q,n}^{\mathrm{corr}}$ in
\eqref{rsd:rejection-rule} preserves the asymptotic
level $\alpha$.
The consistency and local power conclusions in
Theorem~\ref{thm:rsd-power}, and the consistency conclusions
in Proposition~\ref{pro:rsd-rotation}, also remain valid
under this replacement.
\end{proposition}

We use the corrected version in the numerical experiments,
rejecting $H_0$ when
\begin{equation}
T_{Q,n}^{\mathrm{corr}}>z_{1-\alpha}
\quad\text{or}\quad
\bm\theta_0^\prime\bm C_n\bm\theta_0<1+s_n.
\label{rsd:corrected-rejection-rule}
\end{equation}

We next characterize power against sequences of alternatives.
Recall that $a_n=\bm\theta_0^\prime\bm\Sigma_n\bm\theta_0$,
and define
\begin{equation}
\mu_n
=\frac{\sqrt n\,q_n(\bm\theta_0)}
{a_n\sqrt{2c_n\gamma_{2,n}}},
\qquad
h_n=\sqrt n(1+s_n-a_n).
\label{rsd:power-shifts}
\end{equation}
The nonnegative quantity $\mu_n$ measures departure
from an eigenvector relation, while $h_n$ is the standardized difference between
the threshold $1+s_n$ and the population value $a_n$.

\begin{theorem}
\label{thm:rsd-power}
Suppose that the conditions of Theorem~\ref{thm:rsd-null}
hold, and consider a sequence of alternatives.

(i) If
\begin{equation}
\max\{\mu_n,h_n\}\longrightarrow\infty,
\label{rsd:consistency-condition}
\end{equation}
then
\[
\mathbb P\!\left\{
T_{Q,n}>z_{1-\alpha}
\ \text{or}\
\bm\theta_0^\prime\bm C_n\bm\theta_0<1+s_n
\right\}\longrightarrow1.
\]

(ii) If
\[
\mu_n\longrightarrow\mu\in[0,\infty),
\qquad h_n\longrightarrow-\infty,
\]
then
\begin{equation}
\mathbb P\!\left\{
T_{Q,n}>z_{1-\alpha}
\ \text{or}\
\bm\theta_0^\prime\bm C_n\bm\theta_0<1+s_n
\right\}
\longrightarrow1-\Phi(z_{1-\alpha}-\mu),
\label{rsd:local-power}
\end{equation}
where $\Phi$ denotes the standard normal distribution function.
\end{theorem}

Theorem~\ref{thm:rsd-power} shows how each rejection
condition contributes to power.
For a direction lying entirely in the background
eigenspace, $q_n(\bm\theta_0)=0$, and consistency follows
from $\bm\theta_0^\prime\bm C_n\bm\theta_0<1+s_n$
holding with probability tending to one.
For mixtures of spike directions with distinct eigenvalues,
$a_n$ remains above the threshold for all sufficiently
large $n$, and departure from $H_0$ is reflected in
$q_n(\bm\theta_0)$ and hence in $T_{Q,n}$.
For directions mixing spike and background components,
both rejection conditions may contribute.

To characterize power more explicitly, we decompose $\bm\theta_0$ into its
orthogonal projections onto the eigenspaces associated
with the distinct eigenvalues of $\bm\Sigma_n$.
For this discussion, let
$\lambda_{1,n},\ldots,\lambda_{L_n,n}$ denote the distinct spike
eigenvalues, with repeated eigenvalues grouped together, and set
$\lambda_{0,n}=\tau_n$. Thus $L_n\le r_n$.
Let $\mathcal E_{j,n}$ and $\bm\Pi_{j,n}$ denote the eigenspace
and orthogonal projector associated with $\lambda_{j,n}$,
respectively. For a deterministic unit vector $\bm\theta_0$, we define
\begin{equation}
t_{j,n}=\|\bm\Pi_{j,n}\bm\theta_0\|,
\qquad j=0,\ldots,L_n,
\qquad \sum_{j=0}^{L_n}t_{j,n}^2=1.
\label{rsd:grouped-coordinates}
\end{equation}
For each nonzero projection, take
$\bm v_{j,n}=\bm\Pi_{j,n}\bm\theta_0/t_{j,n}$;
when $t_{j,n}=0$, choose any unit vector in $\mathcal E_{j,n}$.
Then
$\bm\theta_0=\sum_{j=0}^{L_n}t_{j,n}\bm v_{j,n}$,
where the $\bm v_{j,n}$ are mutually orthogonal.
Thus each eigenspace contributes one coordinate to the
representation of the given direction, irrespective of its
dimension.

\begin{proposition}
\label{pro:rsd-rotation}
Suppose that the conditions of Theorem~\ref{thm:rsd-null} hold.
For the coordinates in (\ref{rsd:grouped-coordinates}),
\begin{equation}
\begin{aligned}
a_n
&=\sum_{j=0}^{L_n}\lambda_{j,n}t_{j,n}^2,\qquad
q_n(\bm\theta_0)
=\sum_{0\le j<k\le L_n}
(\lambda_{j,n}-\lambda_{k,n})^2t_{j,n}^2t_{k,n}^2.
\end{aligned}
\label{rsd:grouped-population-quantities}
\end{equation}

(i) The quantities in (\ref{rsd:power-shifts}) can be written as
\begin{equation}
\begin{aligned}
\mu_n
&=\frac{\displaystyle\sqrt n
\sum_{0\le j<k\le L_n}
(\lambda_{j,n}-\lambda_{k,n})^2t_{j,n}^2t_{k,n}^2}
{\displaystyle
\left(\sum_{j=0}^{L_n}\lambda_{j,n}t_{j,n}^2\right)
\sqrt{2c_n\gamma_{2,n}}},\qquad
h_n
=\sqrt n\left(1+s_n-
\sum_{j=0}^{L_n}\lambda_{j,n}t_{j,n}^2\right).
\end{aligned}
\label{rsd:grouped-power-shifts}
\end{equation}
If
\begin{equation}
\max\left\{
\sqrt n\sum_{0\le j<k\le L_n}
(\lambda_{j,n}-\lambda_{k,n})^2t_{j,n}^2t_{k,n}^2,\;
h_n
\right\}\longrightarrow\infty,
\label{rsd:grouped-exact-consistency}
\end{equation}
then
\[
\mathbb P\!\left\{
T_{Q,n}>z_{1-\alpha}
\ \text{or}\
\bm\theta_0^\prime\bm C_n\bm\theta_0<1+s_n
\right\}\longrightarrow1.
\]
If instead $\mu_n\to\mu\in[0,\infty)$ and
$h_n\to-\infty$, then
\begin{equation}
\mathbb P\!\left\{
T_{Q,n}>z_{1-\alpha}
\ \text{or}\
\bm\theta_0^\prime\bm C_n\bm\theta_0<1+s_n
\right\}
\longrightarrow1-\Phi(z_{1-\alpha}-\mu).
\label{rsd:grouped-local-power}
\end{equation}

(ii) Define
\begin{equation}
g_n=\min_{0\le j<k\le L_n}
|\lambda_{j,n}-\lambda_{k,n}|,
\qquad
A_n=1-\sum_{j=0}^{L_n}t_{j,n}^4.
\label{rsd:grouped-gap-dispersion}
\end{equation}
Then $q_n(\bm\theta_0)\ge g_n^2A_n/2$.
The test in (\ref{rsd:rejection-rule}) is consistent if
\begin{equation}
\max\left\{
\sqrt n\,g_n^2A_n,\;h_n
\right\}\longrightarrow\infty.
\label{rsd:grouped-consistency}
\end{equation}
In particular, if $\liminf_{n\to\infty}A_n>0$,
then $n^{1/4}g_n\to\infty$ is sufficient for consistency.
\end{proposition}

\begin{remark}
\label{rem:rsd-angle-gap}
Under the conditions of Proposition~\ref{pro:rsd-rotation},
suppose that, for some fixed $\vartheta_*>0$,
\[
\min_{0\le j\le L_n}\arccos(t_{j,n})
\ge\vartheta_*,
\qquad g_n\asymp\delta_n.
\]
Thus the candidate direction stays a positive angle away
from every eigenspace, and the smallest eigenvalue gap
is of the same order as the spike increments.
By (\ref{rsd:grouped-population-quantities}) and
(\ref{rsd:signal-condition}),
\[
q_n(\bm\theta_0)\asymp\delta_n^2,
\qquad
\mu_n\asymp\sqrt n\,\delta_n^2
=(n^{1/4}\delta_n)^2.
\]
\end{remark}

\subsection{Simulation}
\label{subsec:rsd-simulation}

We examine the finite-sample size and power of RSD with
$n=1000$, $p\in\{500,1000\}$ and a nominal significance
level of $0.05$.
We generate independent observations from a multivariate
$t_4$ distribution:
\begin{equation}
\bm x_i=\frac{\bm T^{1/2}\bm g_i}{\sqrt{w_i/4}},
\qquad
\bm g_i\sim N_p(\bm 0,\bm I_p),
\qquad
w_i\sim\chi^2_4,
\label{eq:rsd-sim-data}
\end{equation}
where $\bm g_i$ and $w_i$ are independent and
$\bm T=\operatorname{diag}(1+d_1,\ldots,1+d_r,1,\ldots,1)$.
The quantities $d_1,\ldots,d_r$ are spike increments
above the unit background in the scatter matrix.
We implement the corrected RSD procedure in
\eqref{rsd:corrected-rejection-rule}, rejecting when
\[
T_{Q,n}^{\mathrm{corr}}>z_{0.95}
\quad\text{or}\quad
\bm\theta_0^\prime\bm C_n\bm\theta_0<1+s_n.
\]
We compare $s_n=n^{-1/4}$ and $s_n=n^{-1/6}$,
both of which satisfy \eqref{rsd:threshold-rate}.
Each experiment uses $2000$ independent replications.

To assess size, we set $r=5$ and
$(d_1,\ldots,d_5)=d(2,1.75,1.5,1.25,1)$ for
$d\in\{0.5,1\}$.
Let $\bm e_j$ denote the $j$th coordinate vector in
$\mathbb R^p$.
We test each of the five null directions
$\bm\theta_0=\bm e_j$, $j=1,\ldots,5$, separately, and
Table~\ref{tab:rsd-size} reports empirical rejection
frequencies.

\begin{table}[!htbp]
\centering
\caption{Empirical size (\%) of corrected RSD at the
nominal level of $5\%$, with $n=1000$.
The spike increments are $d(2,1.75,1.5,1.25,1)$.
Each entry is based on $2000$ replications.
Within each $(p,d)$, both thresholds $s_n$
and all five null directions use the same samples.}
\label{tab:rsd-size}
\begin{tabular}{rrr@{\qquad}rrrrr}
\hline
 & & & \multicolumn{5}{c}{Null direction $\bm\theta_0$} \\
$p$ & $d$ & $s_n$
& $\bm e_1$ & $\bm e_2$ & $\bm e_3$ & $\bm e_4$ & $\bm e_5$ \\
\hline
500 & 0.5 & $n^{-1/4}$ & 4.80 & 4.40 & 5.55 & 5.25 & 4.65 \\
500 & 0.5 & $n^{-1/6}$ & 4.80 & 4.40 & 5.55 & 5.25 & 5.10 \\
500 & 1 & $n^{-1/4}$ & 5.55 & 4.85 & 4.95 & 5.35 & 4.95 \\
500 & 1 & $n^{-1/6}$ & 5.55 & 4.85 & 4.95 & 5.35 & 4.95 \\
\addlinespace
1000 & 0.5 & $n^{-1/4}$ & 5.10 & 5.40 & 4.70 & 4.90 & 5.05 \\
1000 & 0.5 & $n^{-1/6}$ & 5.10 & 5.40 & 4.70 & 4.90 & 5.35 \\
1000 & 1 & $n^{-1/4}$ & 5.30 & 5.25 & 4.50 & 5.90 & 6.15 \\
1000 & 1 & $n^{-1/6}$ & 5.30 & 5.25 & 4.50 & 5.90 & 6.15 \\
\hline
\end{tabular}
\end{table}
\FloatBarrier

For power, we take $r=4$ with spike increments
$(d_1,d_2,d_3,d_4)=(1,3,5,7)$ and consider three
families of candidate directions:
\begin{align}
\bm\theta_0^{(1)}(\varphi)
&=\cos(\varphi)\bm e_1+\sin(\varphi)\bm e_5,
\notag\\
\bm\theta_0^{(2)}(\varphi)
&=\cos(\varphi)\bm e_2+\sin(\varphi)\bm e_5,
\notag\\
\bm\theta_0^{(3)}(\varphi)
&=\cos(\varphi)\bm e_4+\sin(\varphi)\bm e_1,
\label{eq:rsd-sim-directions}
\end{align}
where $\varphi\in\{0^\circ,5^\circ,\ldots,90^\circ\}$.
The first two rotate the spike directions corresponding
to the smallest and second-smallest spike eigenvalues,
respectively, towards a background direction.
The third rotates the strongest spike direction towards
the weakest one.

\begin{figure}[H]
\centering
\includegraphics[width=\linewidth]
{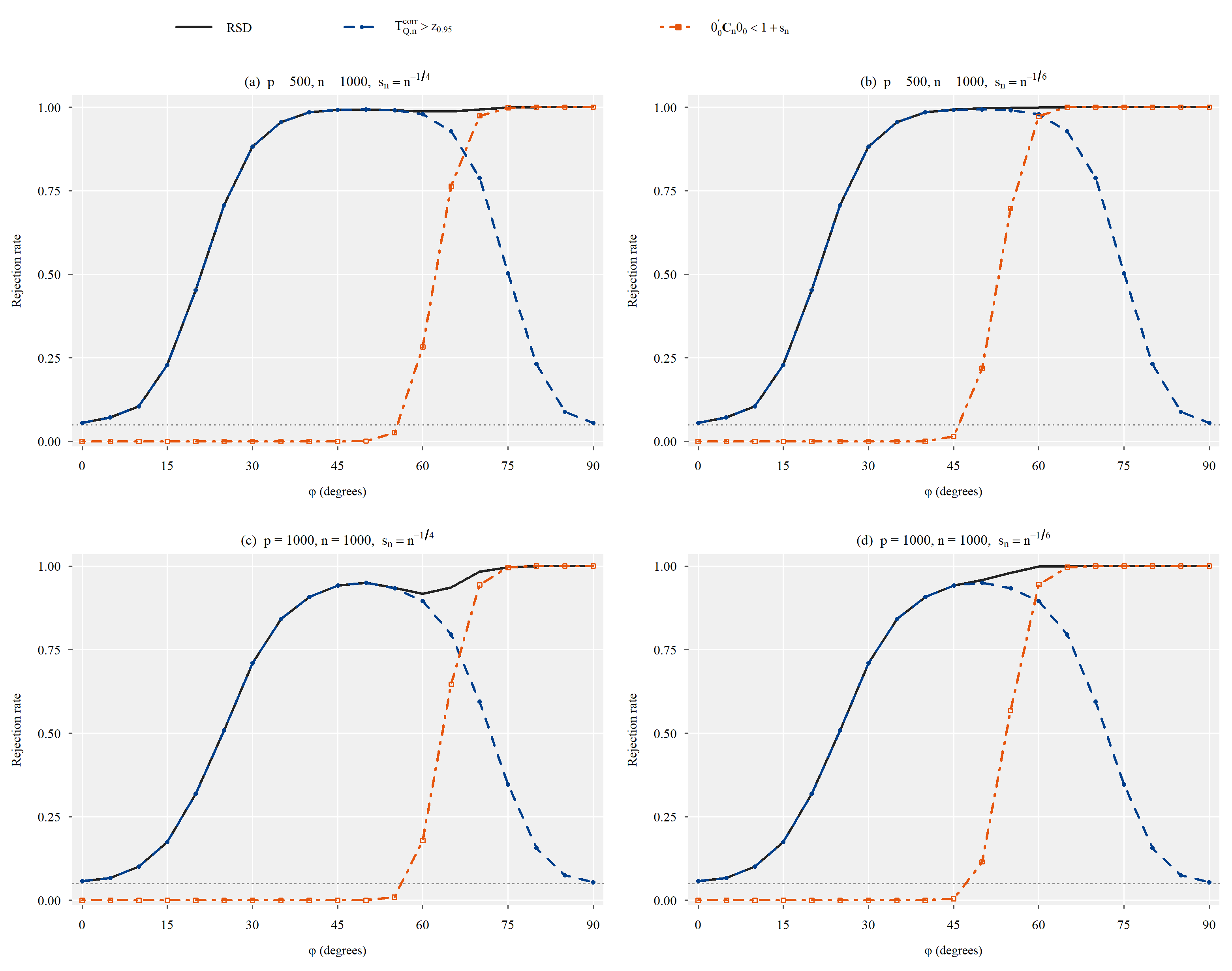}
\caption{RSD rejection rates for rotation from the
smallest spike direction towards the background,
$\bm\theta_0^{(1)}(\varphi)$, with $n=1000$.
Rows: $p=500,1000$; columns: $s_n=n^{-1/4},n^{-1/6}$.
Each rate is based on $2000$ replications.}
\label{fig:rsd-power}
\end{figure}

\begin{figure}[H]
\centering
\includegraphics[width=\linewidth]
{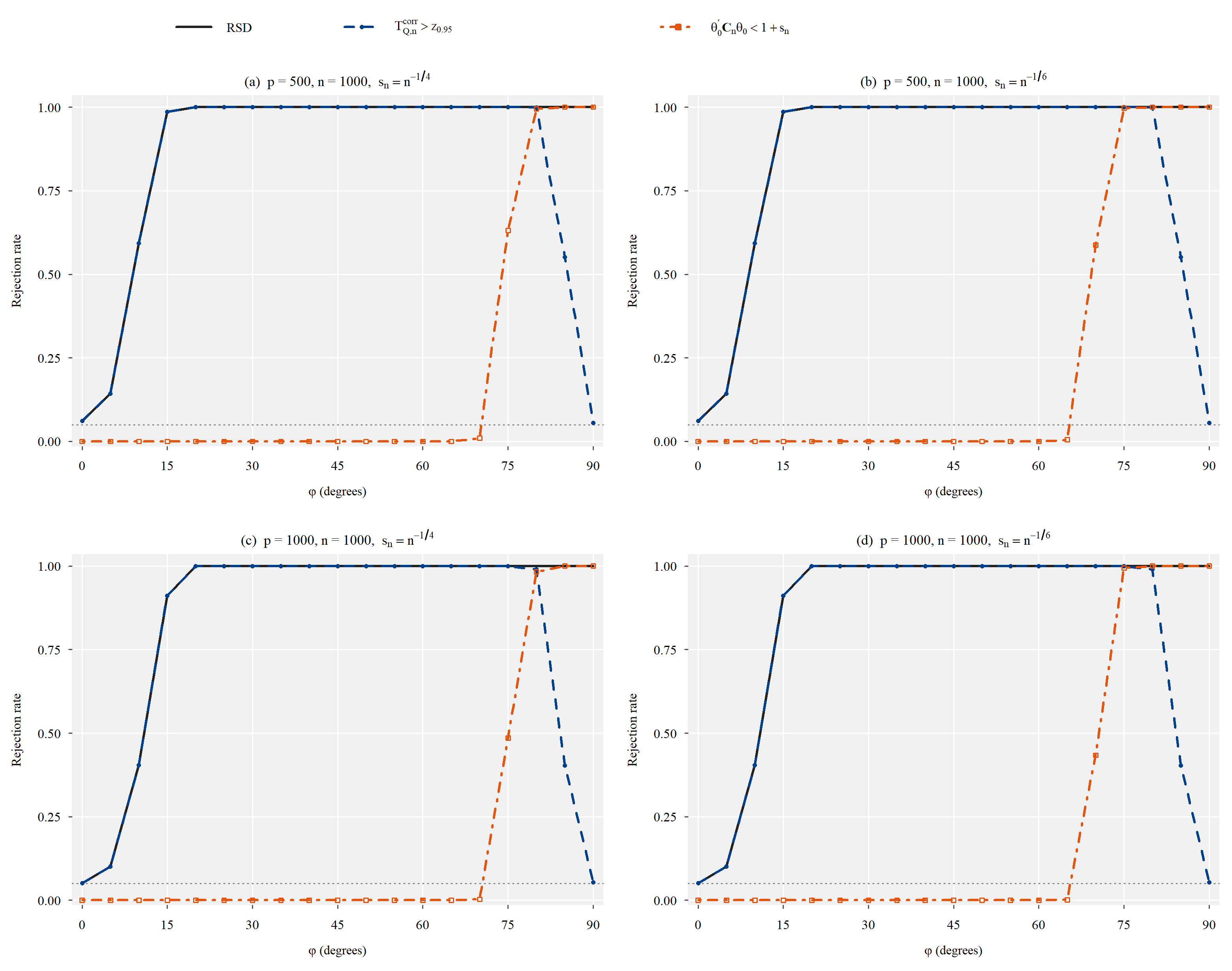}
\caption{RSD rejection rates for rotation from the
second-smallest spike direction towards the background,
$\bm\theta_0^{(2)}(\varphi)$, with $n=1000$.
Rows: $p=500,1000$; columns: $s_n=n^{-1/4},n^{-1/6}$.
Each rate is based on $2000$ replications.}
\label{fig:rsd-power-second}
\end{figure}

\begin{figure}[H]
\centering
\includegraphics[width=0.9\linewidth]
{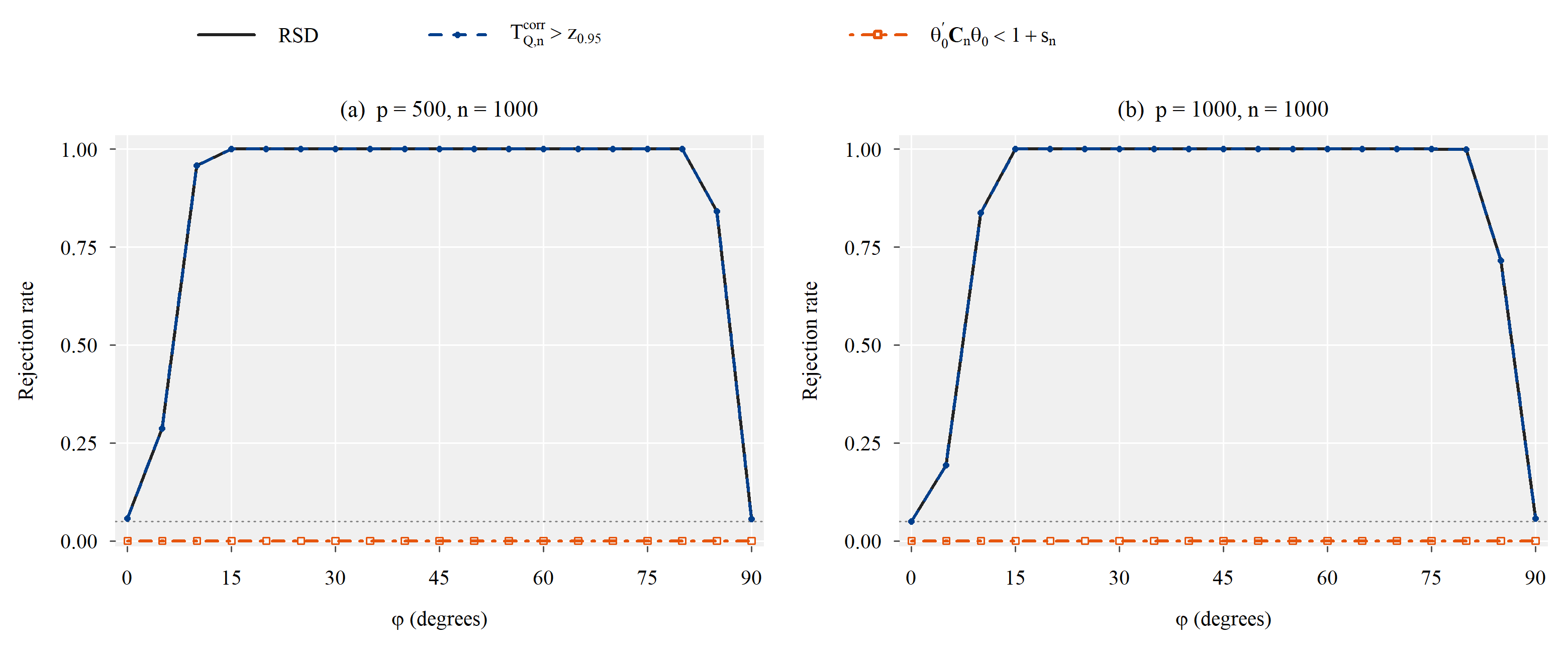}
\caption{RSD rejection rates for rotation between the
strongest and weakest spike directions,
$\bm\theta_0^{(3)}(\varphi)$:
$p=500$ (left) and $p=1000$ (right), with $n=1000$.
Both choices of $s_n$ yield identical empirical rejection
curves. Both endpoints are null configurations.
Each rate is based on $2000$ replications.}
\label{fig:rsd-power-spikes}
\end{figure}

To distinguish the contributions of the two rejection
conditions, we report the empirical frequencies of
$T_{Q,n}^{\mathrm{corr}}>z_{0.95}$ and
$\bm\theta_0^\prime\bm C_n\bm\theta_0<1+s_n$
separately, together with the overall RSD rejection frequency.
The latter is the frequency of the union of the two events,
with simultaneous rejections counted only once.
In Figures~\ref{fig:rsd-power}--\ref{fig:rsd-power-spikes},
black solid curves represent RSD rejection frequencies,
blue dashed curves represent the frequencies of
$T_{Q,n}^{\mathrm{corr}}>z_{0.95}$, and orange dot-dashed
curves represent the frequencies of
$\bm\theta_0^\prime\bm C_n\bm\theta_0<1+s_n$.
The dotted horizontal line marks the nominal level $0.05$.

Figure~\ref{fig:rsd-power} shows rotation from the
smallest spike direction towards the background.
At small and moderate angles, rejection is primarily
due to $T_{Q,n}^{\mathrm{corr}}>z_{0.95}$.
As the direction approaches the background eigenspace,
the condition $\bm\theta_0^\prime\bm C_n\bm\theta_0<1+s_n$
accounts for an increasing proportion of rejections.
With $s_n=n^{-1/4}$, a decrease in overall power remains
visible in the transition region, particularly when $p=1000$.
The larger value $s_n=n^{-1/6}$ raises the threshold and
allows the second rejection condition to contribute at
smaller angles. As a result, the drop in
overall power in the transition region is less pronounced. For $p=n=1000$ and $\varphi=60^\circ$, the condition
$\bm\theta_0^\prime\bm C_n\bm\theta_0<1+s_n$
is satisfied in $17.90\%$ of replications when $s_n=n^{-1/4}$,
compared with $94.45\%$ when $s_n=n^{-1/6}$.
The rejection rate based on $T_{Q,n}^{\mathrm{corr}}>z_{0.95}$
alone remains $89.50\%$ under both choices.
Combining the two rejection conditions, the empirical power
of RSD is $91.65\%$ for $s_n=n^{-1/4}$ and $99.80\%$
for $s_n=n^{-1/6}$.

\FloatBarrier

Figure~\ref{fig:rsd-power-second} shows the corresponding
rotation from the second-smallest spike direction,
with all other settings unchanged.
The larger gap from the background leads to substantially
higher power at small rotation angles.
At $\varphi=15^\circ$, the overall rejection frequencies
are $98.60\%$ for $p=500$ and $91.05\%$ for $p=1000$;
at $20^\circ$, they are $100\%$ and $99.90\%$, respectively.
At every angle from $25^\circ$ to $90^\circ$ on the grid,
all replications are rejected in both dimensions.

\FloatBarrier

Figure~\ref{fig:rsd-power-spikes} shows rotation between
the strongest and weakest spike directions.
The condition $\bm\theta_0^\prime\bm C_n\bm\theta_0<1+s_n$
does not hold in any replication at any angle under
either threshold.
All rejections therefore arise from
$T_{Q,n}^{\mathrm{corr}}>z_{0.95}$, and the two choices
of $s_n$ give identical results.
The empirical power is at least $99.9\%$ at every
angle from $15^\circ$ to $80^\circ$ on the simulation
grid in both dimensions.
These results illustrate the complementary roles
of the two rejection conditions:
$T_{Q,n}^{\mathrm{corr}}$ tests whether $\bm\theta_0$ is an eigenvector, whereas comparing
$\bm\theta_0^\prime\bm C_n\bm\theta_0$ with $1+s_n$
provides power against directions approaching or
lying entirely in the background eigenspace.

\FloatBarrier
\section{\textbf{Real data analysis}}
\label{sec:locata}

We apply RFP and RSD to recording~$1$ of Task~$1$ from the
LOCATA development dataset \citep{evers2020locata}. The recording
contains a single stationary loudspeaker observed by a
$32$-channel Eigenmike spherical microphone array in a reverberant room. The source
and array positions were measured using the OptiTrack optical
tracking system. These measurements provide the geometry for
constructing a candidate signal space.

Following \citet{nakadai2018music}, we construct a steering
vector from the measured source--microphone geometry under a
free-field propagation model. We estimate a whitening
transformation from noise-only portions of the recording and
apply the same transformation to the complex snapshots and
the steering vector. Under the signal-plus-noise model,
whitening by the population noise covariance matrix transforms
the covariance matrix into the sum of an identity matrix and
a low-rank signal component \citep{nadakuditi2010fundamental}.

At each frequency, let $\bm w_j\in\mathbb C^{32}$ denote the
centered and whitened complex snapshot. Under the working
model, we denote
\[
\bm w_j=\zeta_j\bm T_{\mathbb C}^{1/2}\bm u_j^{\mathbb C},
\]
where $\zeta_j>0$ is independent of $\bm u_j^{\mathbb C}$,
which is uniformly distributed on the unit sphere in
$\mathbb C^{32}$ \citep{ollila2012complex}. The Hermitian scatter matrix
$\bm T_{\mathbb C}$ has a common background eigenvalue
$\sigma^2>0$ and a low-rank positive semidefinite perturbation.
We stack the real and imaginary parts \citep{mahot2013asymptotic} to obtain
\[
\bm x_j=
\begin{pmatrix}
\operatorname{Re}\bm w_j\\
\operatorname{Im}\bm w_j
\end{pmatrix}
\in\mathbb R^{64},
\qquad j=1,\ldots,103,
\]
which follows the real elliptical model in \eqref{gabaji}
with scatter matrix
\[
\bm T=
\begin{pmatrix}
\operatorname{Re}\bm T_{\mathbb C}&-\operatorname{Im}\bm T_{\mathbb C}\\
\operatorname{Im}\bm T_{\mathbb C}&\operatorname{Re}\bm T_{\mathbb C}
\end{pmatrix}.
\]
Each eigenvalue of $\bm T_{\mathbb C}$ is preserved in
$\bm T$ with twice its multiplicity. Thus, the background
level and spike increments are preserved, while the dimension
of the spiked eigenspace is doubled. The resulting model
therefore has the structure in \eqref{intro:spiked-scatter}.

We form the spatial signs $\bm y_j=\sqrt p\,\bm x_j/\|\bm x_j\|$
and $\bm C_n=n^{-1}\sum_{j=1}^n\bm y_j\bm y_j'$.
Thus, $p=64$ and $n=103$, with $p/n=0.621$.
Stacking the real and imaginary parts doubles the dimension
but leaves the sample size unchanged.

Let $\bm b\in\mathbb C^{32}$ be the whitened physical steering
vector, normalized so that $\|\bm b\|=1$. Define the two real
candidate directions
\[
\bm\theta_0=
\begin{pmatrix}
\operatorname{Re}\bm b\\
\operatorname{Im}\bm b
\end{pmatrix},
\qquad
\bm\eta_0=
\begin{pmatrix}
-\operatorname{Im}\bm b\\
\operatorname{Re}\bm b
\end{pmatrix}.
\]
These two vectors are orthonormal vectors in $\mathbb R^{64}$, representing
the complex vectors $\bm b$ and $i\bm b$, respectively.
The complex one-dimensional space spanned by $\bm b$ therefore
corresponds to the real two-dimensional space spanned by
$\bm\theta_0$ and $\bm\eta_0$. Its orthogonal projection is
\[
\bm U_0=(\bm\theta_0,\bm\eta_0),
\qquad
\bm{\mathcal Z}_0=\bm U_0\bm U_0'
=\bm\theta_0\bm\theta_0'+\bm\eta_0\bm\eta_0'.
\]
The candidate space has rank $r=2$, and $r/n=0.0194$.

We first apply RFP with $f(z)=z^5$ and $f(z)=z^6$ to test
whether this candidate space equals the population spiked
eigenspace. We consider three frequencies selected through
an exploratory examination of the spectrum.
Table~\ref{tab:locata-rfp} reports the standardized statistics.

\begin{table}[!htbp]
\centering
\caption{Standardized RFP statistics for the LOCATA Eigenmike
recording, with $p=64$, $n=103$ and candidate rank $r=2$.}
\label{tab:locata-rfp}
\begin{tabular}{rrr}
\hline
Frequency (kHz) & $f(z)=z^5$ & $f(z)=z^6$\\
\hline
$3.094$ & $43.02$ & $39.02$\\
$7.078$ & $30.83$ & $29.72$\\
$7.688$ & $32.42$ & $31.20$\\
\hline
\end{tabular}
\end{table}
\FloatBarrier

For both choices of $f$, RFP rejects the null hypothesis at
the nominal level $0.05$ at all three frequencies. Under the
working model, these results provide evidence that the
geometry-based free-field space differs from the population
signal eigenspace.

We next apply RSD to the candidate direction $\bm\theta_0$.
The block structure of $\bm T$ gives, for any
$\lambda>\sigma^2$,
\[
\bm T_{\mathbb C}\bm b=\lambda\bm b
\quad\Longleftrightarrow\quad
\bm T\bm\theta_0=\lambda\bm\theta_0
\quad\Longleftrightarrow\quad
\bm T\bm\eta_0=\lambda\bm\eta_0.
\]
These equivalences follow by comparing the real and imaginary parts and using $T_{\mathbb C}(ib)=iT_{\mathbb C}b$.
Thus, the RSD null hypothesis in \eqref{intro:rsd} for
$\bm\theta_0$ is equivalent to the specified complex
steering direction $\bm b$ being a spike eigenvector.
The two real directions are either both spike eigenvectors
or neither, so it suffices to test $\bm\theta_0$.

Using the same spatial signs and the same three frequencies,
we implement the corrected RSD procedure in
\eqref{rsd:corrected-rejection-rule} at the nominal level $0.05$.
Table~\ref{tab:locata-rsd} reports the corrected statistics and
the corresponding directional quadratic forms.
The two choices $s_n=n^{-1/4}$ and $s_n=n^{-1/6}$ give
thresholds $1+s_n=1.314$ and $1.462$, respectively.

\begin{table}[!htbp]
\centering
\caption{Corrected RSD statistics and directional quadratic
forms for the LOCATA Eigenmike recording, with $p=64$ and
$n=103$. The tested direction $\bm\theta_0$ is the first
column of $\bm U_0$.}
\label{tab:locata-rsd}
\begin{tabular}{rrr}
\hline
Frequency (kHz) & $T_{Q,n}^{\mathrm{corr}}$
& $\bm\theta_0'\bm C_n\bm\theta_0$\\
\hline
$3.094$ & $7.875$ & $3.933$\\
$7.078$ & $10.860$ & $6.158$\\
$7.688$ & $9.024$ & $5.379$\\
\hline
\end{tabular}
\end{table}
\FloatBarrier

At all three frequencies, $T_{Q,n}^{\mathrm{corr}}$ exceeds
the one-sided critical value $z_{0.95}=1.645$, while
$\bm\theta_0'\bm C_n\bm\theta_0$ exceeds both thresholds.
Under the working model,  RSD provides evidence that the specified steering direction is not a spike eigenvector under the working model
while RFP rejects the corresponding candidate signal space.
These findings are consistent with the steering-vector
mismatch discussed by \citet{nakadai2018music}.
\bibliographystyle{elsarticle-harv}
\bibliography{ref}
\end{document}